\def\IEEE{} 

\ifdefined\IEEE
    \documentclass[conference]{IEEEtran}
    \IEEEoverridecommandlockouts
\else
    \documentclass[letterpaper,twocolumn,10pt]{article}
    \usepackage{USENIX}
\fi

\usepackage{cite}

\usepackage{amsmath,amssymb,amsfonts}
\usepackage{algorithmic}
\usepackage{booktabs}
\usepackage{graphicx}
\usepackage{textcomp}
\usepackage{xcolor}
\usepackage{multirow}
\usepackage{colortbl}
\usepackage{hyperref}
\hypersetup{colorlinks=true, unicode=true, linkcolor=[rgb]{0.10,0.05,0.67}, citecolor=[rgb]{0.10,0.05,0.67}, filecolor=[rgb]{0.10,0.05,0.67}, urlcolor=[rgb]{0.10,0.05,0.67}}
\usepackage{tablefootnote}
\usepackage{tikz}
\usetikzlibrary{decorations.pathreplacing}

\begin{document}


\ifdefined\IEEE 

\title    {EPOCH: 
\underline{E}nabling
\underline{P}reemption
\underline{O}peration for
\underline{C}ontext Saving in  
\underline{H}eterogeneous FPGA Systems
}
\else

\title    {\vspace*{-7em}EPOCH: 
\underline{E}nabling
\underline{P}reemption
\underline{O}peration for
\underline{C}ontext Saving in  
\underline{H}eterogeneous FPGA Systems\vspace*{0em}
}
\fi




\author{

\IEEEauthorblockN{Arsalan Ali Malik, Emre Karabulut, and Aydin Aysu
\IEEEauthorblockA{\{aamalik3, ekarabu, aaysu\}@ncsu.edu\\
Department of Electrical and Computer Engineering\\
North Carolina State University\\
\vspace*{-2.0em}
}
}
}



\maketitle
\thispagestyle{plain} 
\pagestyle{plain}     


\ifdefined\IEEE 
    \begin{abstract} 
\else 
\begin{tikzpicture}[remember picture,overlay]

     \draw [decoration={brace,amplitude=0.5em},decorate,ultra thick,black]    
    node[midway,xshift=3.4cm,yshift=2.5cm,right=-5mm,font=\large,text width=5cm]
    {\textbf{Abstract}};
\end{tikzpicture}
    \vspace{-5em}
    \newline
    \indent
\fi
FPGAs are becoming increasingly prevalent in multi-tenant cloud environments. In such an environment, FPGAs are often employed to offload compute-intensive tasks from the main CPU. The operating system (OS) plays a vital role in identifying tasks suitable for offloading and managing the coordination between the CPU and FPGA for seamless task execution. While the OS has long supported preemption as a mechanism for task pause and resume to enhance efficiency and balance CPU time, preempting tasks running on FPGA without losing context poses a significant challenge. Despite the increasing dependence on FPGAs in these dynamic computing environments, FPGA vendors have yet to provide a comprehensive out-of-the-box solution for preemption that considers the full context of tasks. \\\indent
In this paper, we present EPOCH, the first\textit{ vendor-independent} comprehensive framework designed to seamlessly preserve the state of tasks running on multi-tenant cloud FPGAs. EPOCH enables interrupting a tenant's execution at any arbitrary clock cycle, captures its state, and saves this `state snapshot' in off-chip memory with fine-grain granularity. Subsequently, when task resumption is required, EPOCH can resume execution from the saved `state snapshot', eliminating the need to restart the task from scratch. EPOCH automates intricate processes, shields users from complexities, and synchronizes all underlying logic in a common clock domain, mitigating timing violations and ensuring seamless handling of interruptions.\\\indent
We thoroughly verified EPOCH using diverse benchmarks, demonstrating the significance of the proposed work in a real-world scenario with minimal area overhead. EPOCH proficiently captures the state of fundamental FPGA elements, such as look-up tables, flip-flops, block--RAMs, and digital signal processing units. On ZynQ-XC$7$Z$020$ SoC, the proposed solution achieves context save and restore operations per frame in $62.2$ and $67.4$µs, respectively.


\ifdefined\IEEE
\end{abstract}
    \begin{IEEEkeywords}
        Multi-tenant FPGAs, Pre-emption, Context save-and-restore, Partial reconfiguration, State-preservation
    \end{IEEEkeywords}
\else
        \ifdefined\IEEE \vspace{-1.75em}  \fi
\begin{figure}
    \centering
    \ifdefined\IEEE  \else \vspace{-6em}   \fi
        \includegraphics[width=0.49\textwidth]{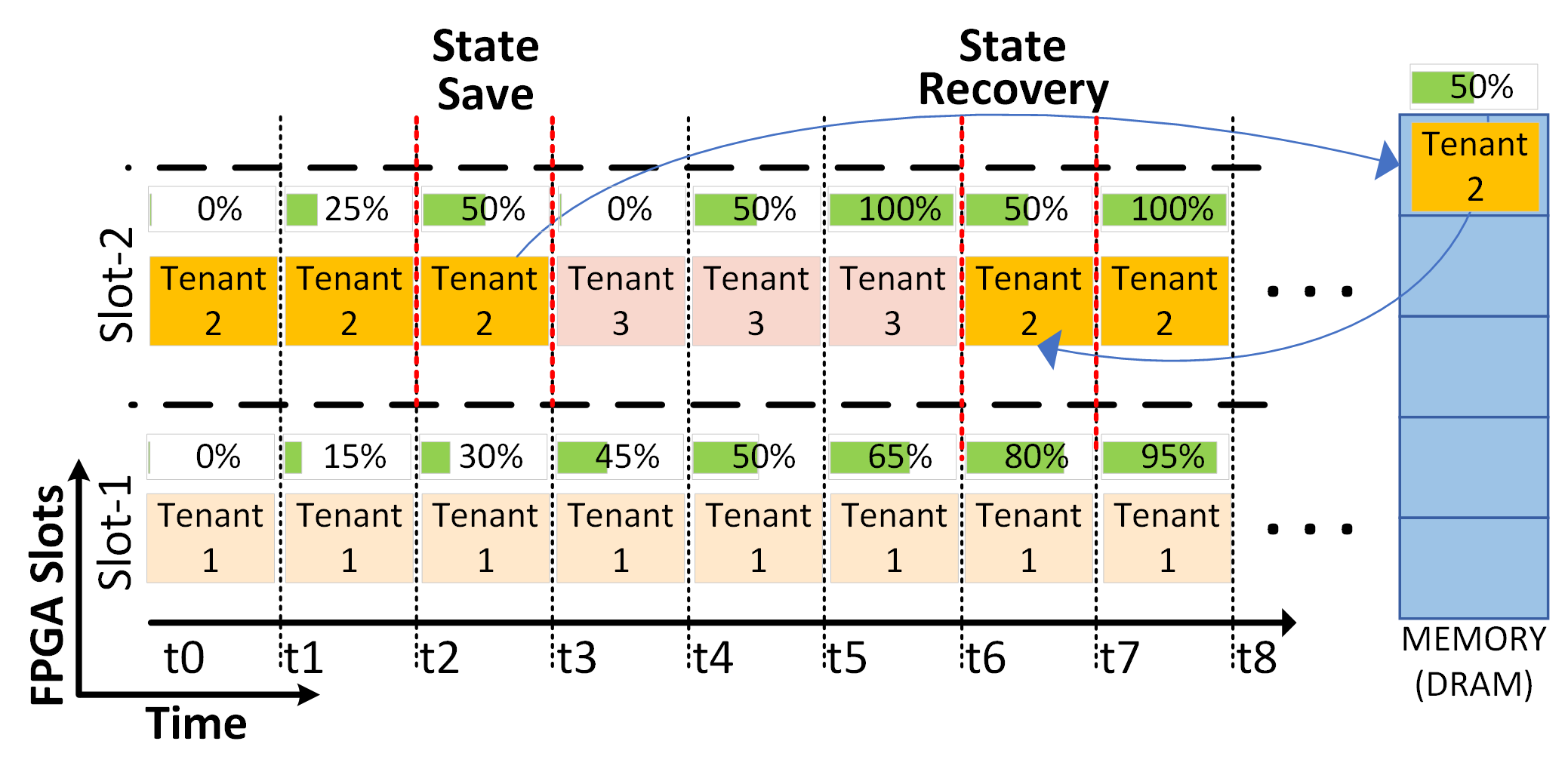}
    \vspace{-1.75em}  
        \caption{An example illustrating preemption support in a multi-tenant FPGA environment. Three tenants share two FPGA slots spatially. At `t3', a higher-priority tenant arrives, prompting the pausing of Tenant-2's work and the preservation of its context. Slot-2 is then allocated to Tenant-3 until completion at `t5'. At `t6', the preserved context of Tenant-2 is restored, enabling a seamless resumption of its execution without any loss of context.}
        \vspace{0.5em}
        \label{fig:pre-emption_Concept}
    \ifdefined\IEEE \vspace{-2em} \else \vspace{-1.75em}  \fi
\end{figure}
\vspace{-0em} 
\fi

\ifdefined\IEEE \vspace{-1.25em}  \else \vspace{-1.0em} \fi
\section{Introduction}
\ifdefined\IEEE \vspace{0em}   \else \vspace{-.5em} \fi
FPGAs have recently entered the cloud environment, marking a significant shift in computing paradigms. The introduction of dynamic partial reconfiguration (PR) has been instrumental in realizing multi-tenancy in the cloud environment~\cite{UG909}. In multi-tenancy, various tenants can share the FPGA fabric either (a) spatially (i.e., the tenants simultaneously occupy distinct physical regions in the layout) or (b) temporally (i.e., the tenants occupy the same physical region in the layout but during different time intervals). 
\\\indent
A tenant can execute one or more tasks in the slot allotted to them to reap the most benefit and achieve higher task utilization during their tenancy\footnote{In this work, the terms ``tasks" and ``tenants" are used interchangeably to refer to the entities involved in the context of FPGA processing and resource allocation in a cloud environment.}. Prior works have proposed mechanisms to enhance the efficiency of FPGA resource sharing through various scheduling algorithms~\cite{STFS, Coyote, AmorphOS, karabulut2024themis}. These scheduling methodologies maintain fairness among tenants while minimizing partial reconfiguration operations. Integrating task preemption with scheduling is a well-established concept within the CPU and the GPU domain. Moreover, leveraging preemption and state/context-saving to switch among tasks or tenants is widely recognized to provide gains in terms of fairness~\cite{ICAPcontext}. 
\\\indent 
Figure~\ref{fig:pre-emption_Concept} exemplifies the use of preemption in scheduling algorithms and its potential advantages in multi-tenant FPGAs. The multi-tenant FPGA is partitioned into two physical slots, each capable of accommodating a maximum of one tenant. For simplicity, Tenant-1, Tenant-2, and Tenant-3 priorities are assumed as `medium', `low,' and `high,' respectively. At time `t0', the scheduling algorithm deploys Tenant-1 to Slot-1 and Tenant-2 to Slot-2. During `t2', a higher-priority tenant requests the FPGA slots. The scheduling algorithm decides to temporarily pause the execution of the task initiated by Tenant-2, given its lower priority. Tenant-2 state is then preserved in off-chip memory, and Slot-2 is allocated to the higher-priority Tenant-3 until its task completes at `t5'. At time `t6,' the preserved state of Tenant-2 is restored, and its execution is resumed on Slot-2.

\ifdefined\IEEE  
\begin{figure}
    \centering
    \ifdefined\IEEE  \else \vspace{-6em}   \fi
        \includegraphics[width=0.49\textwidth]{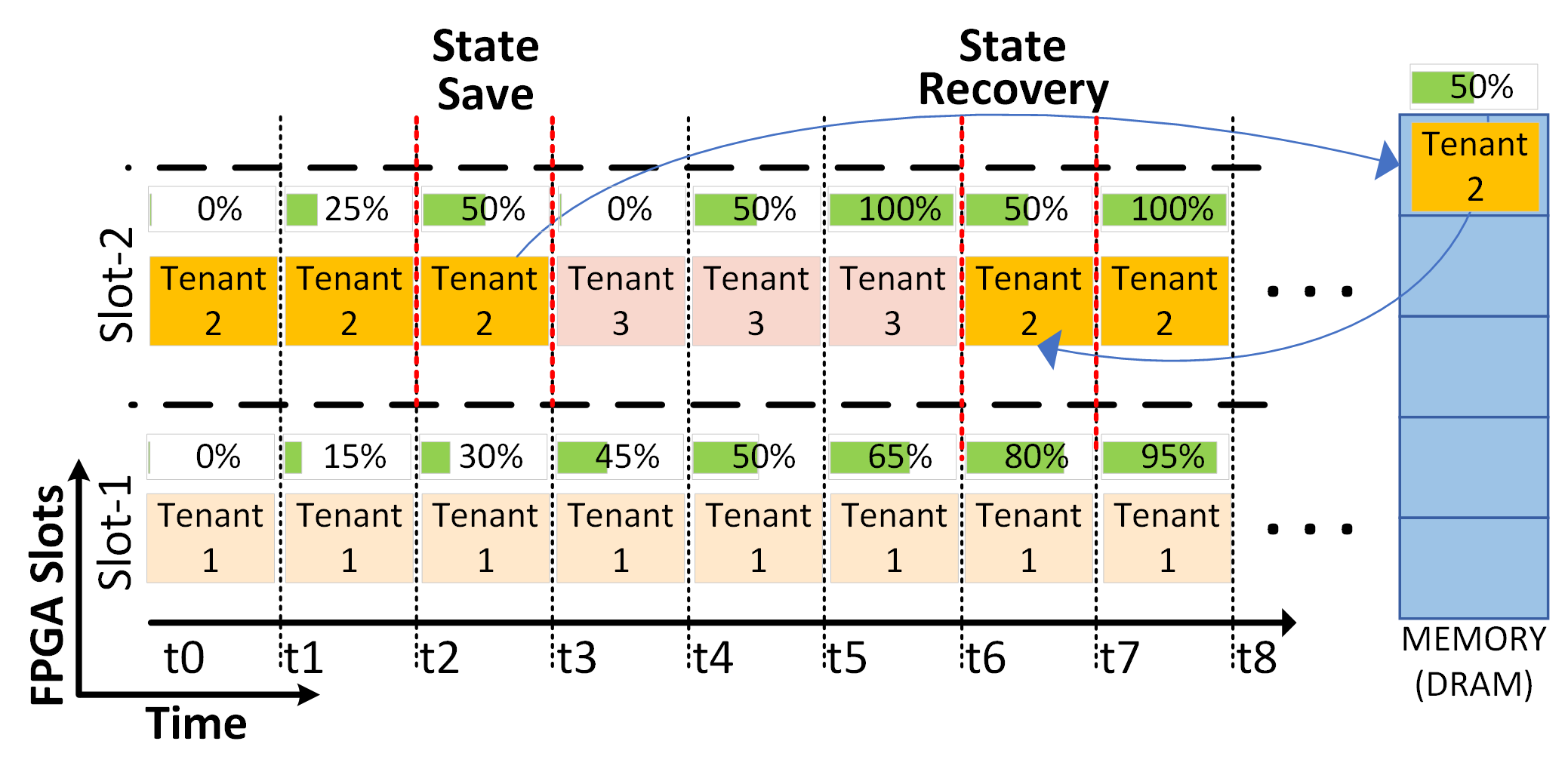}
    \vspace{-1.5em}  
        \caption{An example illustrating preemption support in a multi-tenant FPGA environment. Three tenants share two FPGA slots spatially. At `t3', a higher-priority tenant arrives, prompting the pausing of Tenant-2's work and the preservation of its context. Slot-2 is then allocated to Tenant-3 until completion at `t5'. At `t6', the preserved context of Tenant-2 is restored, enabling a seamless resumption of its execution without any loss of context.}
        \label{fig:pre-emption_Concept}
    \ifdefined\IEEE \vspace{-2em} \else \vspace{-1.75em}  \fi
\end{figure}
\vspace{-0em} 

\vspace{1.5em} \else \vspace{.1em} \fi

\noindent
\textbf{Challenges in FPGA Preemption.} \hspace{0.3em} Achieving preemption on multi-tenant FPGAs is a complex task for three primary reasons: (i) FPGAs consist of various resources, such as look-up tables (LUTs), flip-flops (FFs), block--RAMs (BRAMs), and digital signal processing (DSPs) units, thereby presenting significant challenges to context preservation. The intricacies of FPGA architecture render traditional context preservation mechanisms designed for other embedded devices, such as GPUs or CPUs, ineffective. (ii) The state of each logic element within an FPGA must be read and stored individually. Additionally, this process involves certain manipulations at the bitstream level. The proprietary nature of FPGA architectural information and the limited user guide information on low-level manipulations add further challenges to the process. (iii)  Pausing a running design amid its execution may give rise to hazards, such as timing violations. In the best-case scenario, this could result in corruption of the currently running design, while in the worst-case scenario, it may lead to severe damage to the FPGA itself. \label{Challenges-in-FPGA-Preemption} 
\\\indent
Due to the above-stated reasons, the prior works either lack support for tenant preemption while preserving contextual information or merely provide a \textit{theoretical} preemption support without accounting for the inherent architectural complexity associated with FPGAs~\cite{OPTIMUS}. In prior academic research, two `state' readout methodologies are prevalent. The first approach entails using the scan-chain-based technique, which incorporates additional circuitry into the hardware design to enable the extraction of context~\cite{HW_Checkpoint}. This approach is time-consuming, vendor support dependent, and requires certain hardware guarantees such as signal integrity, isolation, bypassing, error detection, and correction. Furthermore, it does not provide a way for `context' restoration.
\\\indent
The second method involves utilizing a configuration readback mechanism that intelligently leverages the vendor-provided configuration interface for downloading the current design~\cite{ICAPcontext, Virtex6_preemptive}. There are also a few drawbacks to this approach: (i) dependency on a configuration interface/vendor tools, (ii) latency overheads, (iii) lack of \textit{built-in} `context' restoration mechanism, and (iv) complexity and resource utilization. This paper focuses on the second method, to support preemption in cloud FPGAs that can save the state of any design while addressing the drawbacks of design corruption or device failure with minimal area utilization and reduced latencies. 
\\\indent
This paper presents EPOCH: a novel flexible mechanism designed to facilitate the extraction of the active configurations of FPGAs, enabling support for preemption in multi-tenant cloud FPGAs. Our proposed work uses bitstream reverse engineering and flexible clock gating to interrupt a tenant's execution at any arbitrary clock cycle, capture its state, and save it in off-chip memory. EPOCH captures the state of each logic element individually, ensuring precise state preservation with fine-grain granularity. Our proposed approach automates intricate processes, shields users from complexities, and synchronizes all underlying logic using the processing system (PS) clock, mitigating timing violations and ensuring seamless handling of interruptions.
We claim the following four contributions:

\begin{itemize}
\ifdefined\IEEE \vspace{-0.30em} \else \vspace{-.5em} \fi

\item \textbf{Zero-area overhead preemption support.} The suggested framework has zero resource overhead (on the reconfigurable fabric), making it a truly tranquil choice for all designs.
\ifdefined\IEEE  \else \vspace{-.5em} \fi
\item \textbf{Bitstream reverse engineering.} We conducted reverse engineering on the Xilinx Zynq bitstream to introduce the requisite support for preemption. We provide the \textit{first} vendor-independent preemption support that can capture and restore the state of any partition block (P-block) (refer Section~\ref{sec: Proposed-Method} and Table~\ref{tab:Frame_Write_Template}, respectively).

\ifdefined\IEEE  \else \vspace{-.5em} \fi
\item \textbf{Quantitative-benchmarking} We provide the \textit{first} benchmark that facilitates preemption for \textit{any} user design within the framework of a multi-tenant FPGA. Our proposed solution is generic and can accommodate a broad spectrum of user designs.
\ifdefined\IEEE  \else \vspace{-.5em} \fi
\item \textbf{Real-world evaluation.}  We analyze our results on the Xilinx Zynq System-on-Chip (SoC) to model the real-world challenges, limitations of preemption, and the associated potential risks. 
\end{itemize}
\textit{Organization.}\hspace{.5em}The paper is structured as follows: Section~\ref{sec: background} discusses EPOCH's motivation and presents the related work. Section~\ref{Preliminaries} provides an in-depth discussion on FPGA architecture and bitstream composition. Section~\ref{sec: Proposed-Method} outlines our preemption methodology.
Section~\ref{sec: results} presents our findings and evaluates the results.
Section~\ref{sec: Use-cases} outlines cautionary notes, potential side effects, and mitigation strategies to enable safe preemption. Section~\ref{sec: Discussions} delves into the current limitations, potential enhancements, and future avenues for advancing this research. Finally, we conclude the paper in Section~\ref{sec: conclusions}.

\begin{figure*}[t]
    \centering
        \includegraphics[width=\textwidth]{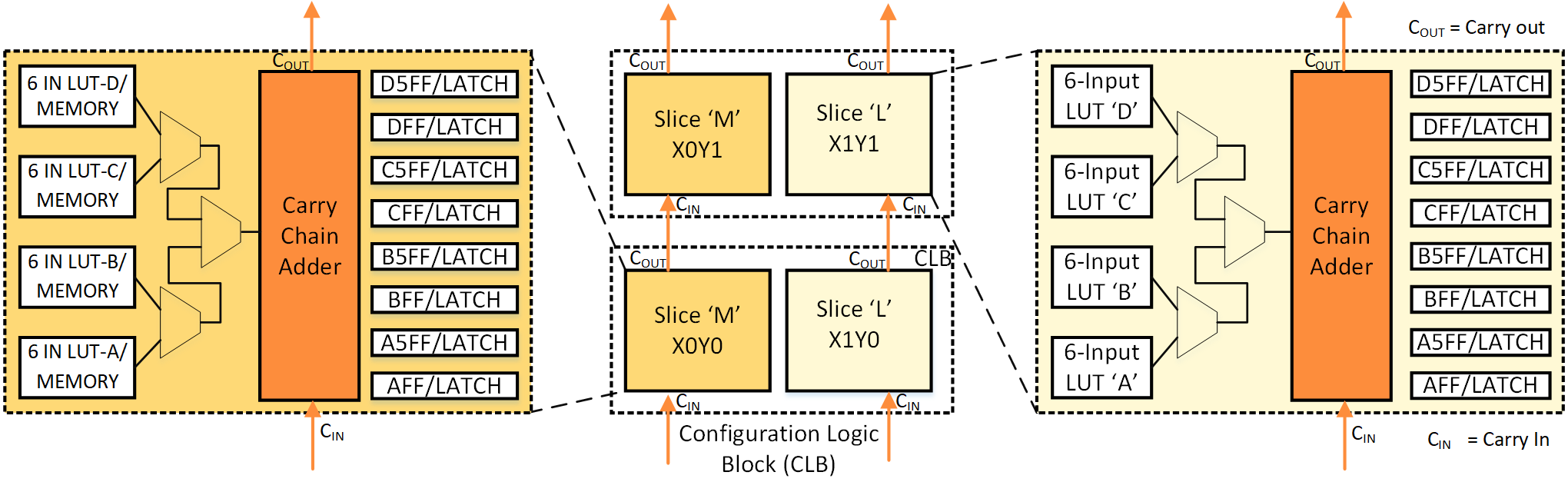}
    \vspace{-2.5em}  
        \caption{The configuration logic block (CLB) of Xilinx $7$-Series FPGA serves as the primary mapping space for user logic. The CLB comprises two slices, Slice-L and Slice-M, each containing four $6$-input look-up tables, eight storage elements (flip-flops/latches), and a carry chain logic. Slice-M can also be utilized as a $32$-bit shift register or to store data via distributed RAM. }
        \label{fig:slice-view}
    \ifdefined\IEEE
        \vspace{-1em} 
    \else
        \vspace{-1.5em} 
    \fi
\end{figure*}

\ifdefined\IEEE
    \vspace{-0em} 
\fi
\section{Motivation and Related Work} \label{sec: background}
\ifdefined\IEEE
    \vspace{0em} 
\fi

Support for preemption has been a subject of investigation for several decades. Substantial research emphasizes the advantages of a successful preemption methodology in scheduling. In recent prior work, OPTIMUS claims to have FPGA preemption support in their scheduling algorithm\cite{OPTIMUS}. Despite their claims, OPTIMUS does not implement FPGA preemption. OPTIMUS's hypervisor only relays the preempt instruction to the accelerator. The guest application (tenant) is responsible for setting up the preemption buffer, checking the task's state, and finishing the work after it has been interrupted.
\\\indent
The problem with this approach is two-fold:  First, it places the entire responsibility of implementing preemption and addressing its associated challenges on the tenant. Second, it assumes that the tenant possesses knowledge of the physical layout of the multi-tenant FPGA and the available underlying resources it is currently occupying. Dario et al.'s work highlights the importance of memory access fairness in cloud FPGAs and promotes preemptive scheduling~\cite{Coyote}. Analogous to prior work~\cite{OPTIMUS}, they assign the responsibility of preemption to the user, necessitating a \textit{user-driven} implementation of preemption, context preservation, and recovery.
\\\indent
In another work, Sameh et al. introduced and later improved support for task interruption and checkpointing~\cite{StateMover,StopnLook}. These works utilize a context-save-and-restore (CSR) database generated during the design synthesis and require integrating several top-level Verilog wrapper functions tailored for context extraction. Despite the minimal overhead associated with these wrappers, their work exhibits the following limitations: (i) To get to the internal state, their work needs to use certain hardware primitives, like virtual input-output (VIO), and is limited to designs that have the AXI/Avalon interfaces. (ii) Their solution is limited to the joint-test action group (JTAG) interface, which is relatively slow. In multi-tenant cloud FPGAs, access to the JTAG interface is generally not provided, thus further limiting the solution to stand-alone FPGAs. (iii) The user must explicitly mark the design or logic that necessitates context saving at \textit{design-time}.
\\\indent
STFS introduces spatial and temporal tenancy in the scheduling algorithm proposed for multi-tenant FPGAs, arguing that this eliminates the necessity for preemption in a multi-tenant environment~\cite{STFS}. The limitation of this approach lies in the utilization of a fixed, small decision interval, which may not be well-suited for the various applications hosted in the cloud. Additionally, the algorithm is based on the unrealistic assumption that multiple partial reconfiguration (PR) slots can be dynamically merged or reduced at runtime to accommodate the area requirements of different tenants. However, it is essential to note that, currently, FPGA vendors do not support such a feature because it causes critical path violations.


\ifdefined\IEEE
    \vspace{0em} 
\fi
\section{Preliminaries}\label{Preliminaries}
\ifdefined\IEEE
    \vspace{0em} 
\fi
Understanding FPGA design and bitstream configuration is essential for a comprehensive grasp of the preemption process. Subsections~\ref{Architecture} and~\ref{Bitstream} provide this information.
\vspace{0em}
\ifdefined\IEEE
    \vspace{0em} 
\fi
\subsection{Granular Architecture and Heterogeneity}\label{Architecture}

Xilinx FPGAs encompass diverse components that facilitate users in developing and updating their designs. 
The building block of a $7$-Series FPGA is known as a configuration logic block (CLB), which houses two slices, namely Slice-L \textit{(Logic)} and Slice-M \textit{(Memory)}, as shown in Figure~\ref{fig:slice-view}. Each slice contains eight storage elements, four of which can be set up as edge-triggered D-type flip-flops or level-sensitive latches, while the remaining four storage components can only be set up as edge-triggered D-type flip-flops. Furthermore, the CLB incorporates four 6-input LUTs and an arithmetic carry chain. The LUT in the Slice-M can also be configured as a $32$-bit shift register or a distributed RAM~\cite{CLB_Guide}. 
\\\indent
An on-chip memory known as block--RAM is provided within an FPGA to accommodate diverse user design configurations with moderate memory storage requirements.
Modern FPGAs also contain DSP slice, a functional block that is optimized for performing mathematical operations commonly used in signal processing tasks\footnote{Although FPGA chips may include other specialized primitive, their usage is highly specific and generally infrequent, thus falling beyond the scope of this research.}.
The elements enumerated above fulfill the requirements of most user designs, where the user-provided logic undergoes a synthesis process, resulting in a netlist. The output of the synthesis process is a bitstream file, encompassing the initial configuration and connection (routing) information. The bitstream is uploaded to the FPGA through a configuration interface configuring the underlying logic elements, such as LUT, FF, BRAM, and DSP. Once configured, the value of each logic element can dynamically change based on the implemented logic. Hence, extracting and preserving the state of these logic elements is crucial to facilitate preemption in an FPGA.

\ifdefined\IEEE
\fi
\subsection{Bitstream Structure and Composition} \label{Bitstream}
\ifdefined\IEEE
    \vspace{0em} 
\fi
The bitstream serves as the configuration data for FPGA configuration memory. Xilinx offers two methods for generating bitstreams as outlined in the configuration guide and a recent study~\cite{Config_guide, stoddard2016hybrid}. First, the *.bit-file contains metadata about the platform it is designed to execute on, the time-stamp, and the actual configuration. Second, the *.bin file contains each FPGA logic element's configuration commands and initial configuration. The bitstream is composed of 4-byte instructions and data words. Downloading this bitstream onto FPGA configuration memory is supported through various interfaces, such as JTAG, Select-MAP, internal-configuration access port (ICAP), and processor-configuration access port (PCAP).

The key distinction among these configuration interfaces lies in the data rate and the bitstream word configuration endianness. Partial reconfiguration introduced the generation of a third type of bitstream known as a partial bitstream~\cite{PR_book}. The primary difference between a full bitstream and a partial bitstream is its size, with the latter being constrained to configure a \textit{specific} region within the FPGA fabric.


The bitstream manipulation has three key components: (i) frame, (ii) frame address register, and (iii) logic location~\cite{BITMAN}. The details of these key components are as follows:
\subsubsection{Frame}
A bitstream file configures the FPGA fabric in small segments known as a `frame'. In the context of $7$-Series FPGAs, a frame consists of $101$ \texttt{words}, with each \texttt{word} being $32$-bits wide~\cite{Config_guide}. Consequently, the minimum size of the bitstream generated by the vendor-provided tool is one frame wide ($3232$-bits). A unique frame address identifies each frame and is addressed by writing into a dedicated configuration register known as the frame address register (FAR).


\subsubsection{Frame Address Register (FAR)}
The FAR register consists of $32$-bits partitioned into the following configurations\footnote{Where the values in square brackets correspond to the bit position within a \texttt{word}~\cite{Config_guide}}.
\begin{itemize}
    \item \textbf{Reserved [$31$--$26$].}  These bits remain unused in $7$-Series devices and are set to zero.
\ifdefined\IEEE     \vspace{.25em} \else \vspace{-.25em} \fi
    \item  \textbf{Block type [$25$--$23$]. }These bits set the target of configuration within the FPGA block. The permissible values are $000$ for CLB blocks, $001$ for BRAM, and $010$ for configuration CLBs (CFG--CLB). In a standard bitstream, only values $000$ and $001$ are present. The inclusion of value $010$ occurs when PR and  \texttt{Reset-after-Reconfiguration} settings are enabled. 
\ifdefined\IEEE     \vspace{.25em} \else \vspace{-1.5em} \fi
    \item \textbf{Top/Bottom row selection [$22$].} This configuration bit selects between the top ($0$) and bottom-half rows ($1$).
\ifdefined\IEEE     \vspace{.25em} \else \vspace{-.25em} \fi
    \item \textbf{Row address [$21$--$17$].} These bits select the current row. The row addresses follow an incrementing pattern to the top in the layout, then reset and increment from the center to the bottom.
\ifdefined\IEEE     \vspace{.25em} \else \vspace{-.25em} \fi
    \item  \textbf{Column address [$16$--$07$.]} These bits determine a resource column, such as a CLB column. Column addresses initiate at zero on the left and increment towards the right (also known as a major column).
\ifdefined\IEEE     \vspace{.25em} \else \vspace{-.25em} \fi
    \item \textbf{Minor address [$6$--$0$].} These bits specify a resource within a major column. \textit{e.g.,} Refer to Figure~\ref{fig:slice-view}, which presents a close-up of a CLB. This CLB consists of Slices `M' and `L'. These slices are numbered and accessed inside the FPGA fabric by manipulating minor address bits.
\end{itemize}    


\ifdefined\IEEE
    \vspace{0em} 
\fi
\subsubsection{Logic Location File}\label{logic-location}
A logic location file is a unique file that has an entry for each context-relevant storing element (LUT, FF, distributed RAM, block--RAM, and DSP) of a design. Utilizing the data contained in the logic location files makes it feasible to ascertain the FAR address of every logic element, thereby enabling the preservation of its state. While producing the *.bit-file, the logic location file (*.ll) can be produced as a secondary product (during the \textit{generate bitstream} step). However, the logic location file does not contain the FAR of the LUTs utilized in the design; therefore, further work is necessary to extract these FARs, as expanded upon in Section~\ref{Complex Benchmarks}.
\begin{figure}[t!]
    \centering

    \includegraphics[width=0.49\textwidth]{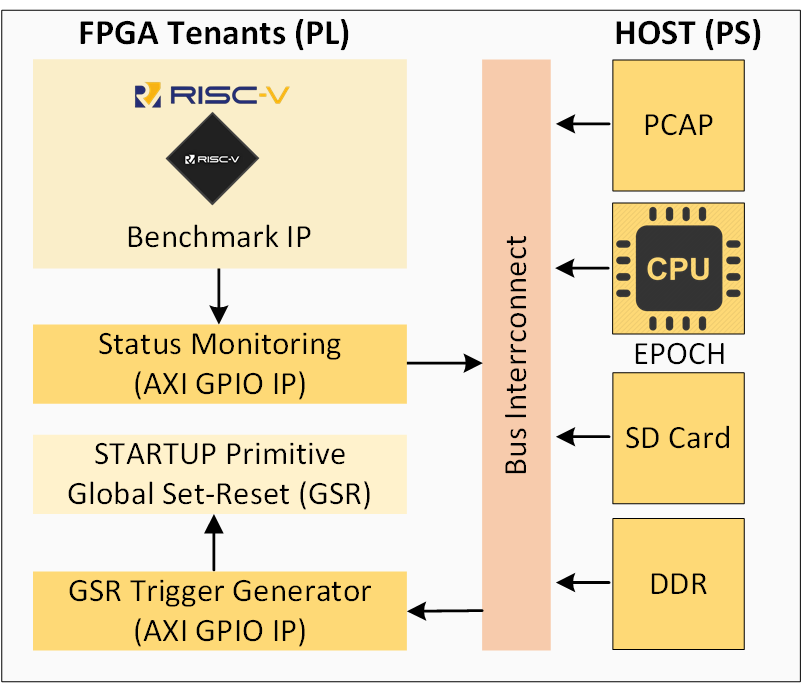}
     \vspace{-1.5em}
         \ifdefined\IEEE     \vspace{-.75em}  \fi
        \caption{Block diagram of EPOCH on the Xilinx Zynq SoC with EPOCH executing on the processing system (PS) and a test/benchmark intellectual property (IP) running on the programmable logic (PL) side. }
        \label{fig:Block Diagram}
\ifdefined\IEEE     \vspace{-1.0em}  \else \vspace{-1.5em} \fi
\end{figure}
\ifdefined\IEEE     \vspace{0em}  \fi
\section{The Proposed Preemption Methodology: EPOCH}\label{sec: Proposed-Method}
\ifdefined\IEEE     \vspace{0em}  \fi
This study presents a framework to support preemption operations for successful context-saving in heterogeneous FPGA systems. 
The proposed methodology consists of four main initiatives. First, it has zero area overhead on the reconfiguration fabric. Second, it provides the \textit{first}-vendor-independent preemption support that captures the state of each logic element in heterogeneous FPGAs. Third, it provides the first generic benchmark that enables preemption operation for \textit{any} user design. Fourth, it models the preemption on Xilinx Zynq SoC to model real-world challenges and the practical limitations.  
\vspace{-0.25em}
\ifdefined\IEEE     \vspace{0em}  \fi
\subsection{State Preservation}
\vspace{0em}
Figure~\ref{fig:Block Diagram} presents the block diagram of our implementation. In an FPGA, LUTs, FFs, BRAMs, and DSPs units are the primary components holding design information. To capture the state of these components, we employed the PCAP interface, exclusive to Xilinx SoCs, offering two key advantages. First, it is a built-in interface on the PS side, eliminating the need for explicit instantiation, unlike ICAP. This simplifies design complexity, particularly for novice users. Second, it reserves no PL resources, resulting in zero area overhead. This feature directly benefits resource-constrained designs, allowing seamless integration at no additional expense. We used the AXI-GPIO\footnote{Advanced eXtensible Interface-General Purpose Input/Output} IP to monitor the status of the benchmark IP and to trigger the \texttt{Global-Set-Reset(GSR)} signal (see Section~\ref{sec: Use-cases} for more details). The attached SD-Card is used to hold the partial bitstreams of each tenant.

Preserving the state of the running design requires pausing its execution, which can be accomplished in two ways: (i) explicit user definition and (ii) clock gating. Explicit user definition requires additional signals/logic to facilitate the state capture functionality. This raises design complexity and necessitates thorough design-logic planning. Conversely, clock gating selectively deactivates (gating) the clock signal to certain circuit elements that are not actively processing data or conducting computations. This involves adding logic (clock buffers) to selectively enable or disable the clock signal to specific circuit parts. Xilinx SoCs provide the implementation flexibility to clock the user design from the PS side. The user's design can, thus, use the PS side clock signal to allow direct control from software, eliminating the need for additional high-performance buffers to achieve clock gating.
EPOCH uses this \textit{flexible} option to reduce design complexity.
\ifdefined\IEEE     \vspace{.25em} \else  \vspace{1em}  \fi

\noindent
\textbf{Mitigation of Challenges in FPGA Preemption.} EPOCH effectively addresses the challenges outlined in Section~\ref{Challenges-in-FPGA-Preemption} through the following three strategies:
\begin{enumerate}
    \ifdefined\IEEE       \vspace{-.15em} \else \vspace{-.25em} \fi
    \item Recognizing the heterogeneous distribution of resources in the FPGA fabric, EPOCH captures the state of each FPGA logic element such as LUT, FF, BRAM, and DSP, individually. This is crucial for preemption in FPGA, given the absence of dedicated registers or memory analogous to CPU stacks\footnote{ In a CPU, stacks store the state of the executing process  (\textit{e.g.}, registers, program counter), facilitating context switching and preemption in multi-tasking environments, ensuring smooth execution and efficient function management.} for preserving the state of ongoing user designs or tenants.
    \ifdefined\IEEE       \vspace{-.15em} \else \vspace{-.25em} \fi
    \item While reading each logic element's state individually incurs some time overhead, it ensures maximum accuracy in state preservation, offering fine-grain granularity. In the case of LUTs and FFs, some bits are masked, which must be unmasked before a successful readback can occur. Likewise, the state of BRAM must undergo certain treatment before it can be restored. EPOCH manages this automatically, shielding users from complex, intricate details.
    \ifdefined\IEEE       \vspace{-.15em} \else \vspace{-.25em} \fi
    \item EPOCH synchronizes all underlying logic on the reconfiguration fabric, including AXI-GPIO, \texttt{GSR}, and the benchmark IP (RISC-V), using the PS-side clock to establish a common clock domain. This common clock mitigates timing violations, enabling the user design to seamlessly handle interruptions and resumptions without introducing timing issues. Moreover, EPOCH leverages \texttt{Reset-after-Reconfiguration (RAR)} and \texttt{GSR} for safe initialization of FF contents, preventing hazards during PR operations. For further insights, refer to Section~\ref{sec: Use-cases}.
    \ifdefined\IEEE       \vspace{-.15em} \else \vspace{-.5em} \fi
\end{enumerate}
\begin{table}
\ifdefined\IEEE     \vspace{-1.9em}   \else \vspace{-5.8em} \fi 
\caption{Our modified command sequence initiates the read-back through the PCAP interface, capturing the complete contents of LUTs, FFs, BRAMs, and DSPs. The `yellow' rows differ (in number) from the command sequence in Table 6-2 of~\cite{Config_guide}, while the `orange' rows depict the proposed modifications to read masked values of LUTs and FFs with zero area overhead.}
\ifdefined\IEEE     \vspace{0em}   \else \vspace{0.5em} \fi 
\label{tab:Frame_Read_Template}

\Huge
\resizebox{\columnwidth}{!}{%
\begin{tabular}{|l|l|l|}
\hline
\textbf{Value} &
  \textbf{Name} &
  \textbf{Purpose} \\ \hline
\cellcolor{yellow!50}\begin{tabular}[c]{@{}l@{}}0xFFFFFFFF\\(8-words)\end{tabular} &
\cellcolor{yellow!50}Dummy Word &
\cellcolor{yellow!50}{-} \\ \hline
0x000000BB &
  \multicolumn{1}{l|}{\begin{tabular}[c]{@{}l@{}}Bus Width \\ Sync Word\end{tabular}}&

   \begin{tabular}[c]{@{}l@{}}  In parallel configuration modes, the bus width\\ selection word is used to automatically set the \\configuration bus width.\end{tabular} \\ \hline
   
0x11220044 &
  \multicolumn{1}{l|}{\begin{tabular}[c]{@{}l@{}}Bus Width \\ Detect \end{tabular}} &
  \begin{tabular}[c]{@{}l@{}}There is a special word called Sync that is used  to\\ensure that the configuration logic lines up with the\\$32$-bit word boundaries.
  \end{tabular} \\ \hline
0xFFFFFFFF &
  Dummy Word &
  \multicolumn{1}{c|}{-} \\ \hline
0xAA995566 &
  \multicolumn{1}{l|}{\begin{tabular}[c]{@{}l@{}}Synchronization\\Word\end{tabular}}&
  \begin{tabular}[c]{@{}l@{}}The FPGA does not process any packets until the Sync\\ word is found. This also serves as a start sequence.\end{tabular} \\ \hline
\cellcolor{yellow!50}\begin{tabular}[c]{@{}l@{}}0x20000000\\(2-words)\end{tabular} &
\cellcolor{yellow!50}NOOP &
\cellcolor{yellow!50}Do not perform any operation. \\ \hline
0x30008001 &
  \multicolumn{1}{l|}{\begin{tabular}[c]{@{}l@{}}Type-1\\Command Word\end{tabular}}&
  \begin{tabular}[c]{@{}l@{}}Writes the command register in the configuration\\ memory to cause the fabric shutdown.\end{tabular} \\ \hline
0x0000000B &
  Fabric shutdown  &
  Shutdown value \\ \hline
\cellcolor{yellow!50}\begin{tabular}[c]{@{}l@{}}0x20000000\\(2-words)\end{tabular} &
\cellcolor{yellow!50}NOOP &
\cellcolor{yellow!50}Do not perform any operation. \\ \hline
0x30008001 &
  \multicolumn{1}{l|}{\begin{tabular}[c]{@{}l@{}}Type-1\\Command Word\end{tabular}}&
    \begin{tabular}[c]{@{}l@{}} Writes to the CRC command register in the \\configuration memory.\end{tabular} \\ \hline

0x00000007 &
    \multicolumn{1}{l|}{\begin{tabular}[c]{@{}l@{}}Reset CRC\\Register\end{tabular}}&
  Resets the internal CRC register. \\ \hline
\begin{tabular}[c]{@{}l@{}}0x20000000\\ (6-words)\end{tabular} &
  NOOP &
  Do not perform any operation. \\ \hline
\cellcolor{orange!50}0x3000C001 &
\multicolumn{1}{l|}{\cellcolor{orange!50}\begin{tabular}[c]{@{}l@{}}Global LUT\\Mask\end{tabular}}&
\cellcolor{orange!50}\begin{tabular}[c]{@{}l@{}}Writes to the CTL0 register to unmask all \\changeable memory cells.\end{tabular} \\ \hline
\cellcolor{orange!50}0x00000100 &
\cellcolor{orange!50}Data Word &
\cellcolor{orange!50}Set GLUTMASK bit in CTL0 register \\ \hline

\cellcolor{orange!50}0x3000A001 &
\multicolumn{1}{l|}{\cellcolor{orange!50}\begin{tabular}[c]{@{}l@{}}Global LUT\\Mask\end{tabular}}&
\cellcolor{orange!50}\begin{tabular}[c]{@{}l@{}}Second-write to the CTL0 register to complete the\\ unmasking.\end{tabular} \\ \hline
\cellcolor{orange!50}0x00000100 &
\cellcolor{orange!50}Data Word &
\cellcolor{orange!50}Set GLUTMASK bit in CTL0 register \\ \hline

\cellcolor{orange!50}0x30008001 &
\multicolumn{1}{l|}{\cellcolor{orange!50}\begin{tabular}[c]{@{}l@{}}Type-1\\Command Word\end{tabular}}&
\cellcolor{orange!50}\begin{tabular}[c]{@{}l@{}}Writes the command register in the configuration \\memory to assert global capture signal for \\ capturing FFs value.\end{tabular} \\ \hline
\cellcolor{orange!50}0x0000000C &
\cellcolor{orange!50}Data Word &
\cellcolor{orange!50}Writes GCAP register \\ \hline
\cellcolor{orange!50}0x20000000 &
\cellcolor{orange!50}NOOP &
\cellcolor{orange!50}Do not perform any operation. \\ \hline
0x30008001 &
  \multicolumn{1}{l|}{\begin{tabular}[c]{@{}l@{}}Type-1\\Command Word\end{tabular}}&
    \begin{tabular}[c]{@{}l@{}}Writes the command register to initiate the\\ fabric readback.\end{tabular} \\ \hline
    
0x00000004 &
  Data Word &
  Read configuration register command \\ \hline
\cellcolor{yellow!50} \begin{tabular}[c]{@{}l@{}}0x20000000\\(3-words)\end{tabular} &
\cellcolor{yellow!50}NOOP &
\cellcolor{yellow!50}Do not perform any operation. \\ \hline
0x30002001 &
  \begin{tabular}[c]{@{}l@{}}Write FAR\end{tabular} &
  Writes the FAR register in the configuration memory \\ \hline
0xXXXXXXXX &
  FAR Address &
  The FAR address whose data is to be read \\ \hline
0x280060CA &
  \begin{tabular}[c]{@{}l@{}}Write FDRO\\Reg Command\end{tabular} &
  Frame data output register command value \\ \hline
0x480000XX &
\begin{tabular}[c]{@{}l@{}}Number of\\Words\end{tabular} &
\begin{tabular}[c]{@{}l@{}}Specifies the number of words that are to be written.\\ The MSB is always set as 0x48, and LSB specifies\\ the number of words to be read.\end{tabular} \\ \hline

\begin{tabular}[c]{@{}l@{}}0x20000000\\ (32-words)\end{tabular} &
  NOOP &
  Do not perform any operation. \\ \hline
\begin{tabular}[c]{@{}l@{}}0xXXXXXXXX\\ (N+101-words)\end{tabular} &
  \begin{tabular}[c]{@{}l@{}}Frame Data\\Read\end{tabular} &
  \begin{tabular}[c]{@{}l@{}}'N' data words being read out from the configuration\\ memory, proceeded by one padding frame of all zeros.\end{tabular} \\ \hline
0x20000000 &
  NOOP &
  Do not perform any operation. \\ \hline

  
0x30008001 &
  \begin{tabular}[c]{@{}l@{}}Write CMD Reg\\ Command\end{tabular} &
  \begin{tabular}[c]{@{}l@{}}Writes the command register in the configuration\\ memory to start the FPGA fabric.\end{tabular} \\ \hline
0x00000005 &
  Start Command &
  Starts the FPGA fabric. \\ \hline
0x20000000 &
  NOOP &
  Do not perform any operation. \\ \hline
0x30008001 &
  \begin{tabular}[c]{@{}l@{}}Write CMD\\ Reg Command\end{tabular} &
  \begin{tabular}[c]{@{}l@{}}Writes the command register in the configuration \\memory to reset CRC value to default.\end{tabular} \\ \hline
0x00000007 &
    \multicolumn{1}{l|}{\begin{tabular}[c]{@{}l@{}}Reset CRC\\Register\end{tabular}}&
  Reset value added in order to bypass CRC check. \\ \hline
0x20000000 &
  NOOP &
  Do not perform any operation. \\ \hline
0x30008001 &
  \multicolumn{1}{l|}{\begin{tabular}[c]{@{}l@{}}Type-1\\Command Word\end{tabular}}&
  \begin{tabular}[c]{@{}l@{}}Writes the command register in the configuration\\ memory to convey that the operation has finished.\end{tabular} \\ \hline
0x0000000D &
  \begin{tabular}[c]{@{}l@{}}De-Synchronization\\ Command\end{tabular} &
  \begin{tabular}[c]{@{}l@{}}No more data is processed after this command is\\ sent. This acts as a write-protection lock.\end{tabular} \\ \hline
\end{tabular}

}
\ifdefined\IEEE     \vspace{-1.0em}   \else \vspace{0em} \fi 
\end{table}


\ifdefined\IEEE    \else \vspace{-2em} \fi
\subsubsection{Context Saving} \label{Context_Saving}
\ifdefined\IEEE    \else \vspace{-.5em} \fi
We propose the following operations for successful context saving and enabling preemption in multi-tenant cloud FPGAs:


\begin{table}[t!]

\ifdefined\IEEE     \vspace{0em} \else  \vspace{-2.5em} \fi
\caption{Our proposed template to write FPGA frames over the PCAP interface. The rows shown in `orange' are specific to the Zynq SoC XC$7$Z$020$-CLG$484$ and must be revised if targeting a different FPGA family.}
\ifdefined\IEEE     \vspace{0em}   \else \vspace{0.5em} \fi 
\label{tab:Frame_Write_Template}
\Large
\resizebox{\columnwidth}{!}{%
\begin{tabular}{|l|l|l|}
\hline
\textbf{Value} &
  \textbf{Name} &

  \textbf{Purpose} \\ \hline

\begin{tabular}[c]{@{}l@{}}0xFFFFFFFF\\ (8-words)\end{tabular} &
  Dummy Word &
 {-} \\ \hline
 0x000000BB &
  \multicolumn{1}{l|}{\begin{tabular}[c]{@{}l@{}}Bus Width \\ Sync Word\end{tabular}}&
   \begin{tabular}[c]{@{}l@{}}  In parallel configuration modes, the bus width\\ selection word is used to automatically set the \\configuration bus width.\end{tabular} \\ \hline
 0x11220044 &
  \multicolumn{1}{l|}{\begin{tabular}[c]{@{}l@{}}Bus Width \\ Detect Word\end{tabular}}&
   \begin{tabular}[c]{@{}l@{}}There is a special word called Sync that is used  to\\ensure that the configuration logic lines up with the\\$32$-bit word boundaries.\end{tabular} \\ \hline
 0xFFFFFFFF &
   Dummy Word &
   {-} \\ \hline
 0xAA995566 &
 \multicolumn{1}{l|}{\begin{tabular}[c]{@{}l@{}}Synchronization\\Word\end{tabular}}&
  \begin{tabular}[c]{@{}l@{}}The FPGA does not process any packets until the \\Sync word is found. This also serves as a start sequence.\end{tabular} \\ \hline
 \begin{tabular}[c]{@{}l@{}}0x20000000\\ (2-words)\end{tabular} &
   NOOP &
   Do not perform any operation. \\ \hline
 0x30008001 &
  \multicolumn{1}{l|}{\begin{tabular}[c]{@{}l@{}}Type-1 \\Command Word\end{tabular}}&
   \begin{tabular}[c]{@{}l@{}}Writes the command register in the configuration \\memory to reset CRC value to default.\end{tabular} \\ \hline
 0x00000007 &
     \multicolumn{1}{l|}{\begin{tabular}[c]{@{}l@{}}Reset CRC\\Register\end{tabular}}&
   Reset value. \\ \hline
 \begin{tabular}[c]{@{}l@{}}0x20000000\\ (2-words)\end{tabular} &
   NOOP &
   Do not perform any operation. \\ \hline
 \cellcolor{orange!50}0x30018001 &
   \cellcolor{orange!50}\begin{tabular}[c]{@{}l@{}}Write IDCODE\\Reg Command\end{tabular} &
   \cellcolor{orange!50}\begin{tabular}[c]{@{}l@{}}Writes the IDCODE command register to specify the \\FPGA the bitstream is intended for.\end{tabular} \\ \hline
  \cellcolor{orange!50}0x03727093 &
  \cellcolor{orange!50}FPGA IDCODE &
  \cellcolor{orange!50}\begin{tabular}[c]{@{}l@{}}Unique IDCODE of FPGA, The shown value belongs\\ to ZynQ SoC FPGA~\cite{Config_guide}\end{tabular} \\ \hline
   0x20000000 &
   NOOP &
   Do not perform any operation. \\ \hline
\cellcolor{orange!50}0x30002001 &
  \cellcolor{orange!50}\begin{tabular}[c]{@{}l@{}}Write FAR \\ Command\end{tabular} &
  \cellcolor{orange!50}Writes the FAR register in the configuration memory \\ \hline
\cellcolor{orange!50}0x004201xx &
   \cellcolor{orange!50}FAR Address &

    \cellcolor{orange!50}\begin{tabular}[c]{@{}l@{}}The FAR address where the data writing is intended.\\ In this example the  FAR 0x004201xx will configure \\the elements of Slice\_X0Y0.\end{tabular} \\ \hline

 0x20000000 &
  NOOP &
  Do not perform any operation. \\ \hline
0x30008001 &
  \begin{tabular}[c]{@{}l@{}}Write CMD\\Reg Command\end{tabular} &
  Prepares the commands register for incoming commands. \\ \hline
0x00000001 &
      \multicolumn{1}{l|}{\begin{tabular}[c]{@{}l@{}}Write\\Configuration Data\end{tabular}}&
  \begin{tabular}[c]{@{}l@{}}Starts transferring the received data of the written to the\\  FPGA configuration memory.\end{tabular} \\ \hline


    

0x20000000 &
  NOOP &
  Do not perform any operation. \\ \hline


0x30004000 &
\begin{tabular}[c]{@{}l@{}}Write FDRI\\Reg Command\end{tabular} &
Frame data input register command value \\ \hline
0x500000CA &
Number of Words &
\begin{tabular}[c]{@{}l@{}}Specifies the number of words that are to be written.\\ The MSB is always set as 0x5 and LSB specifies the\\ number of data plus padding words.\end{tabular} \\ \hline
\begin{tabular}[c]{@{}l@{}}0x00000000\\ (101-Words)\end{tabular} &
Frame Words &
Frame data to be written \\ \hline
\begin{tabular}[c]{@{}l@{}}0x00000000\\ (101-Words)\end{tabular} &
Padding Words &
\begin{tabular}[c]{@{}l@{}}The padding words must follow the frame data.\end{tabular} \\ \hline

     


0x30008001 &
\begin{tabular}[c]{@{}l@{}}Write CMD\\Reg Command\end{tabular} &
\begin{tabular}[c]{@{}l@{}}Writes the command register in the configuration\\ memory to reset CRC value to default.\end{tabular} \\ \hline
0x00000007 &
\multicolumn{1}{l|}{\begin{tabular}[c]{@{}l@{}}Reset CRC\\Register\end{tabular}}&
  Reset value added to bypass CRC check. \\ \hline

\begin{tabular}[c]{@{}l@{}}0x20000000\\ (2-words)\end{tabular} &
  NOOP &
  Do not perform any operation. \\ \hline
\cellcolor{orange!50}0x30002001 &
\cellcolor{orange!50}\begin{tabular}[c]{@{}l@{}}Write FAR \\ Command Word\end{tabular} &
\cellcolor{orange!50}Writes the FAR register in the configuration memory \\ \hline
\cellcolor{orange!50}0x004201xx &
\multicolumn{1}{l|}{\cellcolor{orange!50}\begin{tabular}[c]{@{}l@{}}Next FAR\\Address\end{tabular}}&
\cellcolor{orange!50}\begin{tabular}[c]{@{}l@{}}The next FAR address to be written after the current\\write operation has finished.\end{tabular} \\ \hline
0x30008001 &
    \multicolumn{1}{l|}{\begin{tabular}[c]{@{}l@{}}Type-1\\Command Word\end{tabular}}&
  \begin{tabular}[c]{@{}l@{}}Writes the command register in the configuration\\memory to reset CRC value to default.\end{tabular} \\ \hline
0x00000007 &
      \multicolumn{1}{l|}{\begin{tabular}[c]{@{}l@{}}Reset CRC\\Register\end{tabular}}&
  Reset value added to bypass CRC check. \\ \hline
\begin{tabular}[c]{@{}l@{}}0x20000000\\ (2-words)\end{tabular} &
  NOOP &
  Do not perform any operation. \\ \hline
0x30008001 &
    \multicolumn{1}{l|}{\begin{tabular}[c]{@{}l@{}}Type-1\\Command Word\end{tabular}}&
  \begin{tabular}[c]{@{}l@{}}Writes the command register in the configuration\\ memory to convey that the operation has finished.\end{tabular} \\ \hline
0x0000000D &
  \begin{tabular}[c]{@{}l@{}}De-Synchronization\\ Command\end{tabular} &
  \begin{tabular}[c]{@{}l@{}}Upon receipt of this command by the FPGA, no further\\ data processing occurs. This acts as a write-protection lock.\end{tabular} \\ \hline
0xFFFFFFFF &
Dummy Word &
{-} \\ \hline
\begin{tabular}[c]{@{}l@{}}0x20000000\\ (2-words)\end{tabular} &
  NOOP &
  Do not perform any operation. \\ \hline

\end{tabular}%
}
\vspace{-1.0em}

\end{table}
\begin{itemize}
    \vspace{0em}
    \item \texttt{\textbf{Clock Disable.}} To capture the \textit{running} state of the design, it is essential to freeze its current state. This ensures that no ongoing transactions are in progress, which could modify the content in memory elements. In our experiments, our design was running using a common PS clock, simplifying the process of halting the clock\footnote{We have also tested and verified our methodology using clock buffers to perform clock gating. We discuss issues about multiple clock domains in Section~\ref{sec: Use-cases}.}. We halted the clock by writing to the throttle count control register to freeze the design's \textit{current} state. This register is part of the write-protected slice control register configuration; hence, it must be unlocked first by writing into the SLCR--UNLOCK register~\cite{ZynQ_TRM}. 
    \begin{figure}[t!]
    \centering
        \includegraphics[scale=0.65]{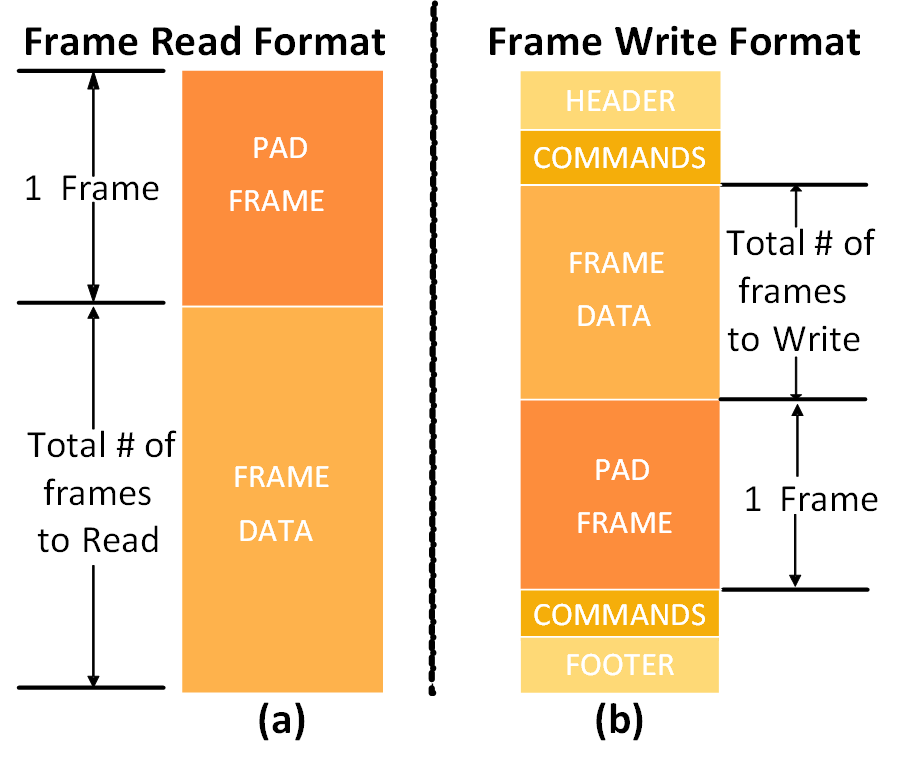}
     \vspace{-1.75em}
        \caption{The frame format for two different configurations: (a) frame readback (context saving) and (b) frame write (context recovery). The position of the padding frame varies during the frame read and write operations. To restore the context, additional header and footer commands must be sent along with the data, as shown in Table~\ref{tab:Frame_Write_Template}. Adding the header and footer makes the state restoration bitstream analogous to a partial bitstream.}
        \label{fig:frame_format}
    \vspace{-2.75em}
\end{figure}
        \ifdefined\IEEE       \vspace{-.15em} \else \vspace{-.5em} \fi
    \item  \texttt{\textbf{Readback-Capture.}} Our modified readback command sequence is elaborated in Table~\ref{tab:Frame_Read_Template}. The rows highlighted in `yellow' are the commands that differ (in terms of number) from those mentioned in Table 6-2 of configuration-guide~\cite{Config_guide}. On the PCAP interface, the readback mechanism fails if the number of commands is inaccurate. The modified command sequence needed to unmask the LUTs and distributed RAM content in the readback data is illustrated in the rows highlighted in `orange'.
    Once the clock is disabled, EPOCH initiates the PL readback mechanism. A basic readback on the PL can only retrieve the content of LUTs. There are two ways to capture the current state of the FFs: (i) incorporate the global capture (GCAP) primitive in the hardware, which also requires the user design to contain the logic that repeatedly triggers the GCAP, or (ii) include the capture command in the readback sequence, which is not subject to this constraint~\cite{Config_guide}.

    EPOCH adopts the latter approach because it reduces design complexity by eliminating the requirement to incorporate the GCAP primitive and its associated trigger logic explicitly. BRAM and unpipelined DSP state can also be captured using the same mechanism. However, the restoration involves some additional steps, as discussed in Section~\ref{Context_Restore}. The capture is executed frame-by-frame, controlled by passing the FAR address of each relevant location intended for state saving. 
    \begin{figure*}
    \centering
          \includegraphics[width=\textwidth]{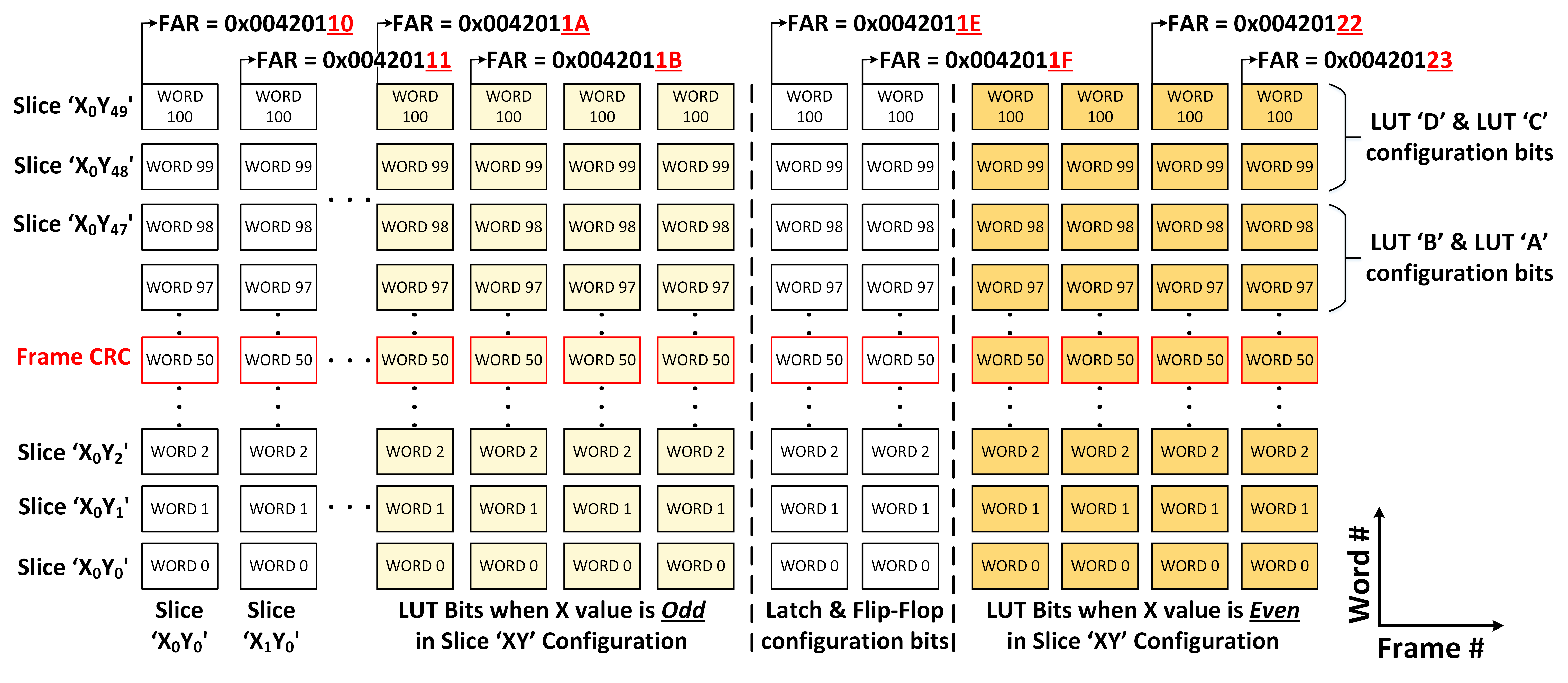}
    \vspace{-2.0em}  
        \caption{Visualization of the frame and configuration bit layout within the Xilinx Zynq $XC7Z020-CLG484$ bitstream. The spatial distribution of \texttt{words} and the varying FAR across the fabric, commencing from Slice\_X$0$Y$0$, is presented for readers' comprehension. The \texttt{word}-$50$ in each frame is reserved for frame CRC. Manipulating the LSBs in the FAR address allows targeting diverse FPGA resources, such as lookup tables or flip-flops.}
        \label{fig:BitstreamArchitecture}
    \vspace{-1em}  
\end{figure*}

        \ifdefined\IEEE       \vspace{-.15em} \else \vspace{-.5em} \fi

    \item  \texttt{\textbf{DRAM Dump.}}
    The bitstream size for each FPGA varies, and the size of the partial bitstream can increase depending on the partition block size. The on-chip BRAMs are not sufficiently large in every FPGA family to adequately hold this data. EPOCH stores the captured data in the off-chip DRAM. When a PL readback is initiated, the user receives one dummy padding frame\footnote{The padding data consists of all zeros and can be discarded.}  followed by the actual frame data, as illustrated in Figure~\ref{fig:frame_format}. The readback data (current state of the user design) can be stored as is but requires further formatting before it can be restored to the PL, as explained in Section~\ref{Context_Restore}.
        \ifdefined\IEEE       \vspace{-.15em} \else \vspace{-.5em} \fi
    \item   \texttt{\textbf{Clock Enable.}} The design-under-test clock is restored to resume its normal operation once the data state has been successfully saved. The specifics are the same as mentioned in the Subsection \texttt{Clock Disable} above.
    
\end{itemize}

        \ifdefined\IEEE       \vspace{0em} \else \vspace{-1em} \fi
\subsubsection{Context Restore} \label{Context_Restore}  
        \ifdefined\IEEE       \vspace{0em} \else \vspace{-.25em} \fi

The initial and final steps for context restoration are identical to those stated in Subsection~\ref{Context_Saving}, namely \texttt{Clock Disable} and \texttt{Clock Enable}. Figure~\ref{fig:frame_format} illustrates the format of the data readback from the PL. This data, obtained over the PCAP interface, consists solely of configuration information and is not replayable as is. For a successful state recovery operation, proprietary configuration commands must proceed and succeed the configuration data exactly as we reverse-engineered and demonstrate in Figure~\ref{fig:frame_format}. The command sequence, which is proprietary and not fully documented in the Xilinx documentation, necessitated reverse engineering efforts. The partial bitstreams generated by Vivado were analyzed to determine the required command sequence. Table~\ref{tab:Frame_Write_Template} lists our proposed template and the complete instruction sequence for a successful state recovery operation. 
A padding frame must be included after the frame data to completely flush the contents to the configuration memory of the PL.
\\\indent
Xilinx incorporates the CRC-32 checksum in the frame data to detect transmission errors. Therefore, at \texttt{word} position 50, each frame write operation must also include a valid 32-bit CRC value\cite{stoddard2016hybrid}. This requirement can be met in three different ways. First, include the logic for calculating the CRC and append it within each frame. Second, modify the constraint file to turn off the CRC check altogether. Third, change how the CRC comparison works by appending the CRC reset command after the frame data. This resets the PL's internally generated CRC to a default value, bypassing the CRC check altogether\cite{stoddard2016hybrid}. We have used the third method for simplicity. However, without the CRC check, there's a risk of undetected transmission errors, leading to a potential misconfiguration of the FPGA.
\\\indent
To target a specific LUT and FF within a frame, one must carefully modify the minor address in a FAR. Figure~\ref{fig:BitstreamArchitecture} presents a visualization of bitstream formation in $7$-Series FPGA, having 6-input LUTs ($2^6$ = $64$ bits for initialization). For the CLB presented in Figure~\ref{fig:slice-view} above, our experiment revealed this CLB's base FAR address to be 0x004201\underline{xx}. To modify the LUTs of Slice `L' within this CLB, the lower bits of FAR (minor address) must have values ranging from 26 to 29, as the $64$-initialization bits are scattered in 4-frames (each holding $16$-bits). Thus, the FAR addresses to target these LUTs can range from 0x004201\underline{1A} to 0x004201\underline{1D} in odd-numbered slices (Slice\_\textbf{X1}Y0). Similarly, for an even-numbered Slice `M' (Slice\_\textbf{X0}Y0), the lower bits of FAR (minor address) must have values ranging from 32 to 35, making the respective FARs 0x004201\underline{20} to 0x004201\underline{23}. The allowable minor values for FF data within these slices are 30 and 31, making the FAR address 0x004201\underline{1E} and 0x004201\underline{1F}\footnote{The minor range values and FAR addresses are specified in decimal and hexadecimal numbering format, respectively. The FAR addresses given in this example are specific to Zynq XC$7$Z$020$-CLG$484$ SoC.}.

\ifdefined\IEEE  \vspace{0em} \else \vspace{-.10em} \fi
The BRAM configuration data requires additional modification for a successful state restoration. In our experiments, we observed that without this BRAM treatment, although the data writing is successful, the PL reverts to the `old-state' it had before initiating the state restoration. A comparison with the originally generated bitstream revealed that BRAM frame readback changes certain bits to `$1$', regardless of whether or not the BRAM content was updated during execution. Consequently, these bits must be reverted to their original value of `$0$' to restore BRAM contents successfully. Equation~\ref{eq_word} specifies the \texttt{word (w)} numbers for a BRAM frame, where bit number $18$ must be modified. The value of \texttt{w} ranges from $4$ to $95$.


\ifdefined\IEEE  \vspace{-1em} \else \vspace{-1.5em} \fi
\begin{equation}
\label{eq_word}
    (\texttt{w} < 54\ \& \ \texttt{w} \% 10 == 4)\ || \ (\texttt{w} > 54\ \& \texttt{w} \ \% 10 == 5)
\end{equation}
\subsubsection{A Real-World Application Scenario}
To gain a deeper understanding of the operational capabilities of EPOCH, let us examine a practical scenario that involves a two-slot configuration running on the shared PS clock (CLK1).
For simplicity, suppose both designs only consist of LUTs and FFs. A 4-bit up-counter is executed in Slot-1, and a 4-bit down-counter is executed in Slot-2; the same signal, \textit{update} influences both counters.  Upon asserting the \textit{update} signal, the up-counter increments by one, and the down-counter decrements by one. The initial values of the counters were `0x0' and `0xF', eventually changing to `0x3' and `0xC' after three assertions of the \textit{update} signal.
\\\indent
At system power-up, EPOCH loads our proposed template of Table~\ref{tab:Frame_Write_Template} in a contiguous manner into DRAM onetime. This loading process populates dummy \textit{frame data} and \textit{FAR} addresses of all zeroes. If a state-saving operation is requested, EPOCH executes four steps. First, it pauses the clock (CLK1) to both slots\footnote{EPOCH operates on a separate CLK0 to prevent design freeze during preemption.}. Second, it initiates the readback capture mechanism for Slot-1. Third, it saves the readback data (frames) in DRAM at a predefined address, such as 0x0000000A. (Repeating steps 2--3 for Slot-2 preserves its state at a different DRAM address, such as 0x000B0000). EPOCH receives $202$-\texttt{words} of data in exchange for a single frame readback. The initial $101$ are discarded as padding \texttt{words} (composed entirely of zeroes); the remaining $101$-\texttt{words} constitute the design `context'. Subsequently, the FAR address information and the `context' data readback from the device are uploaded to the DRAM in place of the initial dummy template data.
\begin{figure}[t!]
    \centering
\includegraphics[width =\columnwidth]{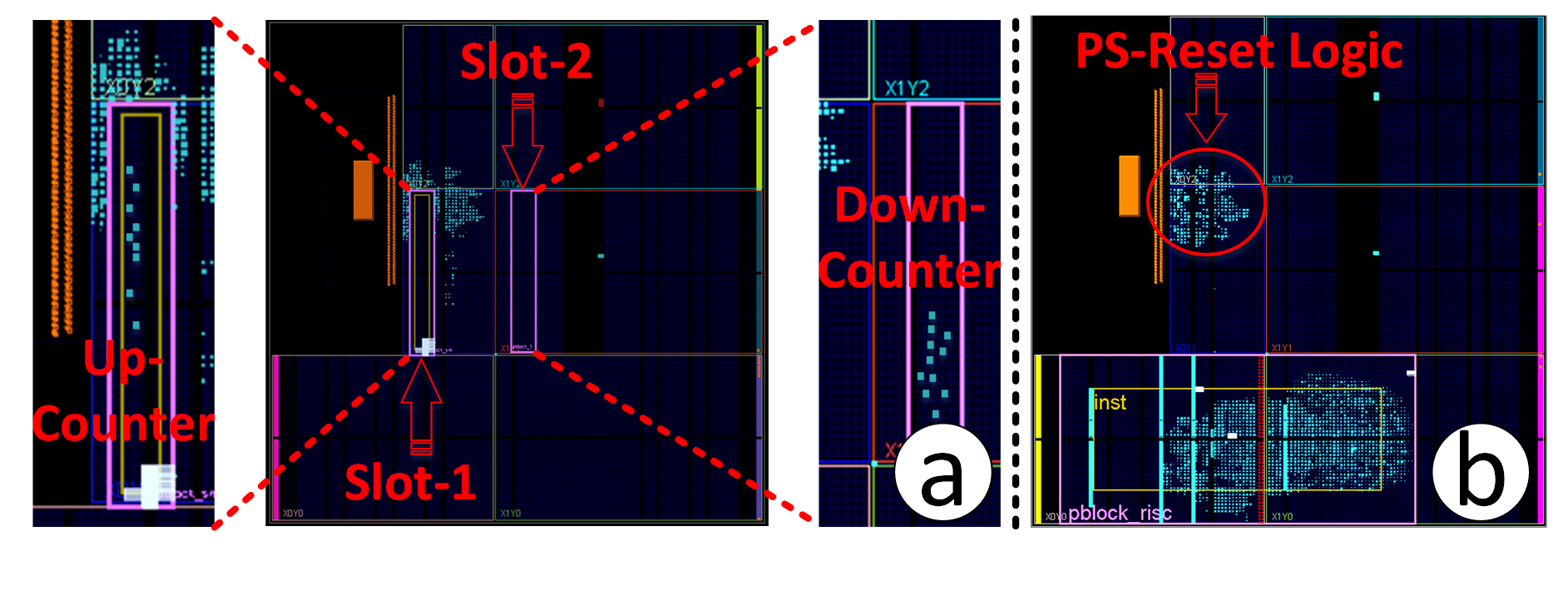}
     \vspace{-1.5em}
        \caption{The two FPGA layouts employed in our experiments for (a) basic and (b) complex benchmarks.  The basic benchmark simulates distinct tenants in a multi-cloud environment and consists of two slots. An example demonstrates the logic of a 4-bit up-counter mapped in Slot-1, while Slot-2 contains the logic mapping of a 4-bit down-counter. In contrast, the complex benchmarks comprise a single slot spanning two clock regions  (X$0$Y$0$ and X$1$Y$0$), encompassing heterogeneous resources, such as LUTs, FFs, BRAMs, and DSP units.}
        \label{fig:FPGA-Layout}
\ifdefined\IEEE  \vspace{-1.4em} \else \vspace{-1.75em} \fi
\end{figure}
In case the readback, data is identified as a BRAM frame\footnote {The third byte of BRAM frames FAR addresses begins with value $0$x$C2$ and is utilized to identify BRAM frames.}, the data is dynamically subjected to the bit-flipping treatment (outlined in Subsection~\ref{Context_Restore}) during the state-saving operation.  Finally, EPOCH restores the clock (CLK1) to both slots, preserving the counters' values as `0x3' and `0xC' in DRAM. Let's consider that after the state saving operation, the \textit{update} signal is asserted numerous times, changing the up and down counters value to `0x7' and `0x8', respectively.

To roll back both counters `state' to the stored values in DRAM, the user reinitiates EPOCH. The restoration also involves four steps. First, it pauses the clock (CLK1) of both designs. Second, it reads DRAM contents (frames) for Slots 1 and 2 from addresses 0x0000000A and 0x000B0000, respectively, followed by initiating the PCAP-write process\footnote{The frames are formatted with header and footer commands on-the-fly as per Table~\ref{tab:Frame_Write_Template} during the readback phase. Writing to PCAP is equivalent to loading a partial bitstream using PCAP. By utilizing our proposed template of Table~\ref{tab:Frame_Write_Template}, the reformatted readback data is essentially a \textit{run-time} generated partial bitstream, having full design context.}. Third, it asserts the \texttt{GSR} signal to transfer values from the bitstream to underlying logic elements like LUTs and FFs. Finally, it resumes the clock (CLK1) to both slots. Upon clock resumption, the state of the up-counter in Slot-1 reverts from `0x7' to `0x3', and in Slot-2, the down counter now has the value of `0xC' instead of `0x8', demonstrating a successful state recovery.

\ifdefined\IEEE \vspace{0em} \else \vspace{-1.35em}  \fi
\section{Results and Comparison} \label{sec: results}
\ifdefined\IEEE      \vspace{0em} \else     \vspace{-.7em} \fi
\textbf{Test-Platform.}  We conducted experiments using the Xilinx Vivado $2018.2$ and Zedboard XC$7$Z$020$ from the Zynq-$7000$ SoC family. Our exploration of preemption in spatial multi-tenancy involved a two-phase approach. In the first phase, we partitioned the PL fabric into two slots and ran applications with smaller area footprints, as shown in Figure~\ref{fig:FPGA-Layout} (a). The purpose of this experiment was design-area exploration and how it affected the preemption. In the second phase, we crafted a versatile benchmark framework that accommodates any user application in a single slot on the PL side, as illustrated in Figure~\ref{fig:FPGA-Layout} (b). This configuration allowed us to pause, preempt, and restore the application's state at any arbitrary clock cycle.


\ifdefined\IEEE      \vspace{-.5em} \else     \vspace{0em} \fi
\subsection{Basic Benchmarks} \label{basic_benchmarks}

\ifdefined\IEEE 
    \vspace{0em}
\else 
    \vspace{-.7em}
\fi
In this series of basic benchmarks, the experiments involved having two partition slots on the PL side. On these slots, we ran dedicated custom hardware applications such as finite state machine (FSM)-based up/down counters that incremented or decremented a $4$-bit register based on an \textit{update} signal. We conducted these experiments at two distinct clock speeds, namely 50 Mhz and 100 Mhz. The \textit{update} signal was controlled by push-button logic. In \textit{Slot-1}, a $4$-bit counter was incremented, while in \textit{Slot-2}, another $4$-bit counter was decremented. After several iterations, the state of both slots was preserved in DRAM, followed by a reset of these slots. The applications from \textit{Slot-1} and \textit{Slot-2} was successfully restored, with the restored slots functioning as intended. 

The same experiment was replicated with two different configurations of the linear feedback shift registers (LFSRs) : (i) $8$-bit and (ii) $32$-bit, each having different seed values. The LFSR changed its value at each rising edge of the clock. The current value of LFSR was routed to the PS for continuous monitoring through the AXI interface. The states of both LFSRs were successfully saved and recovered without breaking/corrupting their functionality. 

Our initial design comprised solely of LUTs and FFs, therefore, the evaluation of preemption effectiveness for BRAM and DSP units remained unexplored. To thoroughly investigate the performance of DSP and BRAM designs, we conducted additional experiments using the SHA-256 cryptographic module to build a hash chain. This design tests the capability of the proposed design to preserve and restore the state of designs involving DSPs and BRAMs. Two SHA-Core implementations were used to construct this hash chain. The experiment encompassed initiating hash computation, pausing the design, saving the state, and proficiently restoring the configuration to resume the hash chain computation. After pausing and saving the state of the hash chain, we systematically re-initialized the slot to its initial blank state before restoring its `context'. This step was undertaken to verify whether the functionality of the hash chain remained intact upon the design's state recovery.
\\\indent
During basic benchmarking, we noticed a need to repeat extensive work for each new design. Moreover, manual parameter adjustments were frequently required for successful state preservation and restoration, including P-block resizing (to map the new application) and identifying FAR addresses. Recognizing these challenges, we created a benchmark framework that streamlines the processes and offers generality for \textit{any} user application. The details of this framework are thoroughly discussed in the subsequent Subsection~\ref{Complex Benchmarks}.

\begin{figure}[t!]
    \centering
    \includegraphics[width=\columnwidth]{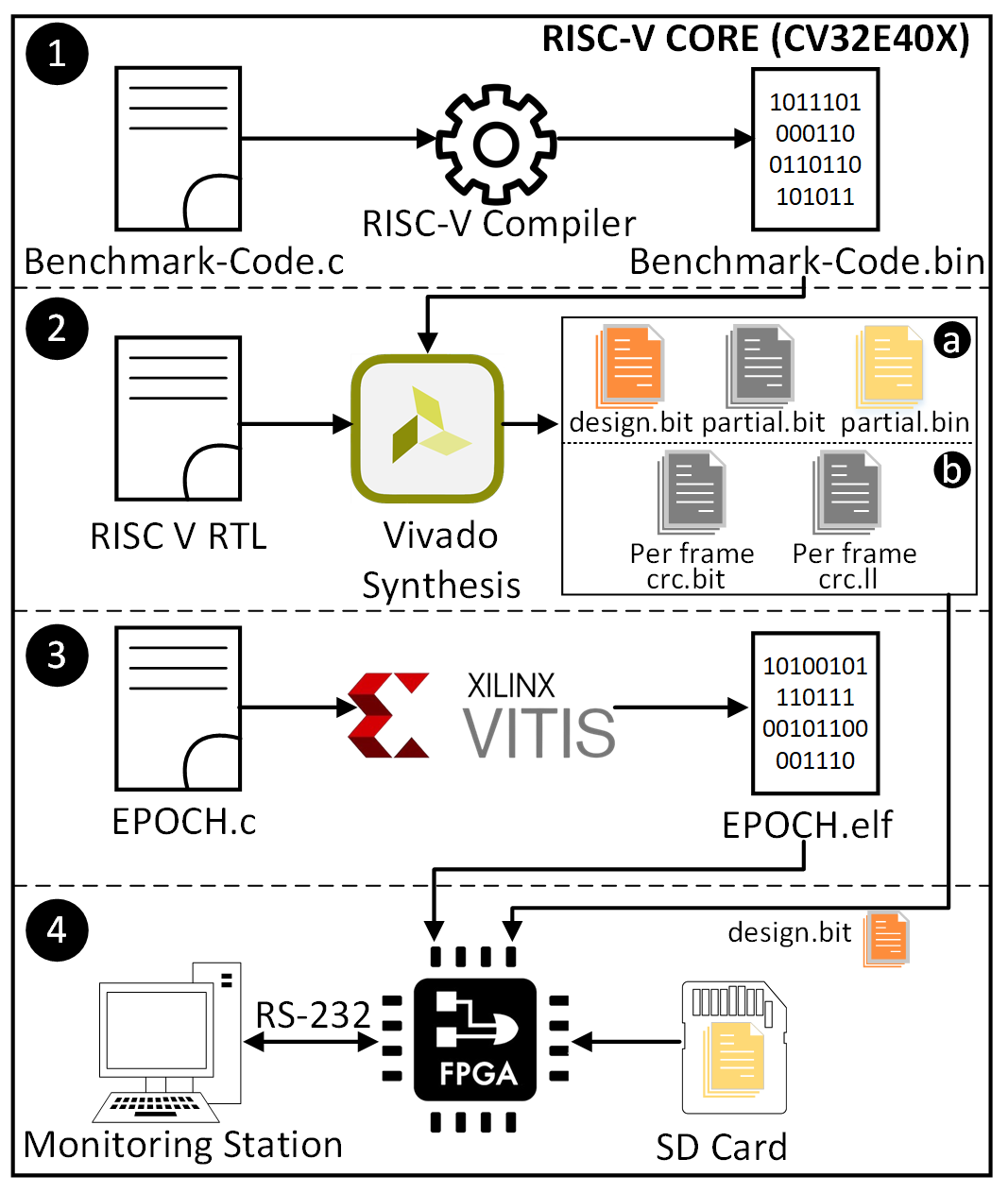}
     \ifdefined\IEEE  \vspace{-2.5em} \else \vspace{-2.5em} \fi
        \caption{The operational workflow of EPOCH consists of four steps. (1) The RISC-V toolchain converts C-based code into a memory initialization file. (2) Using Xilinx Vivado, the synthesis process combines this initialization file with RISC-V RTL code for CV32E40X, resulting in equivalent full-and-partial design bitstreams. This step is iterated with the per-frame CRC setting enabled to generate the FAR address of each logic element used in a design. (3) The EPOCH preemption code runs on the PS side, written in C, using the Xilinx SDK/Vitis platform. (4) The partial binary files are stored on the SD card, and the FPGA is programmed with the full design bitstream and the EPOCH preemption code. FPGA and PC communication is established using USB-to-serial for IP status reporting and monitoring.}
        \label{fig:EPOCH-Process-Flow}
  \ifdefined\IEEE  \vspace{-1.0em} \else \vspace{-2em} \fi
\end{figure}
\ifdefined\IEEE     \vspace{0em} \else \vspace{-1em} \fi
\subsection{Complex Benchmarks} \label{Complex Benchmarks}
\ifdefined\IEEE     \vspace{0em} \else \vspace{-.5em} \fi
We integrated the EPOCH framework with the RISC-V GNU toolchain to generalize the EPOCH framework and make it compatible with running \textit{any} user application. In our experiments, we utilized the CV32E40X RISC-V soft processor core running at 100 MHz. The operational workflow of EPOCH is depicted in Figure~\ref{fig:EPOCH-Process-Flow}. Here are the key steps:
\begin{table*}[t!]
\caption{Resource utilization, latency, and context save/recovery times for complex benchmarks.}
\label{tab:Complex_benchmar_resource}
\centering
\resizebox{\textwidth}{!}{%
\begin{tabular}{|c|c|c|c|c|c|c|c|}
\hline
\textbf{Classification} &
  \textbf{\begin{tabular}[c]{@{}c@{}}Benchmark Name\end{tabular}} &
  \textbf{\begin{tabular}[c]{@{}c@{}}Clock Cycles\end{tabular}} &
  \textbf{LUTs} &
  \textbf{FFs} &
  \textbf{BRAMs} &
  \textbf{DSPs} &
  \textbf{\begin{tabular}[c]{@{}c@{}}Context Save/Recovery Time (ms)\end{tabular}} \\ \hline
\multirow{5}{*}{Cryptography} &
  AES-128 &
  8195 &
  \textbf{3978} &
  \textbf{2353} &
  32 &
  3 &
  \textbf{3.18}/0.020 \\ \cline{2-8} 
 &
  SHA-256 &
  5330 &
  4070 &
  \textbf{2323} &
  32 &
  3 &
  3.17/0.020 \\ \cline{2-8} 
 &
  \begin{tabular}[c]{@{}c@{}}FALCON Key \\ Pair Generate\end{tabular} &
  25551 &
  \multirow{3}{*}{4070} &
  \multirow{3}{*}{\textbf{2356}} &
  \multirow{3}{*}{32} &
  \multirow{3}{*}{3} &
  \multirow{3}{*}{%
    \begin{tabular}[c]{@{}c@{}}3.11/0.019\\ \footnotesize{(Simulation-based results)}\end{tabular}%
  } \\ \cline{2-3}
 &
  \begin{tabular}[c]{@{}c@{}}FALCON Sign. \\ Generate\end{tabular} &
  264590955 &
   &
   &
   &
   &
   \\ \cline{2-3}
 &
  \begin{tabular}[c]{@{}c@{}}FALCON Sign. \\ Verify\end{tabular} &
  706555 &
   &
   &
   &
   &
   \\ \hline
\begin{tabular}[c]{@{}c@{}}Machine\\ Learning\end{tabular} &
  GEMM &
  2681 &
  4056 &
  696 &
  32 &
  3 &
  3.12/0.021 \\ \hline
\multirow{2}{*}{\begin{tabular}[c]{@{}c@{}}Database\\ Operations\end{tabular}} &
  BFS &
  13595 &
  4059 &
  700 &
  32 &
  3 &
  3.17/\textbf{0.026} \\ \cline{2-8} 
 &
  Sort &
  5468 &
  4053 &
  680 &
  32 &
  3 &
  3.15/\textbf{0.019} \\ \hline
\multirow{2}{*}{\begin{tabular}[c]{@{}c@{}}Pattern\\ Recognition\end{tabular}} &
  NW &
  44707 &
  4057 &
  685 &
  32 &
  3 &
  \textbf{3.07}/0.020 \\ \cline{2-8} 
 &
  KMP &
  18935 &
  4053 &
  695 &
  32 &
  3 &
  3.10/0.019 \\ \hline
\multirow{2}{*}{\begin{tabular}[c]{@{}c@{}}Computational\\ Benchmarks\end{tabular}} &
  Dhrystone &
  165207 &
  4055 &
  686 &
  32 &
  3 &
  3.14/0.022 \\ \cline{2-8} 
 &
  \begin{tabular}[c]{@{}c@{}}RISC-V\\ Coremark\end{tabular} &
  2159022 &
  4052 &
  694 &
  32 &
  3 &
  3.11/0.021 \\ \hline
\end{tabular}%
}
\vspace{-1.5em}
\end{table*}

\begin{enumerate}
\ifdefined\IEEE  \else    \vspace{-.250em}  \fi
    \item We employed open-source C codes and compiled them using the RISC-V GNU toolchain within a Linux environment~\cite{RISCV, benchmark, benchmark2, benchmark3, benchmark4}. This toolchain leverages open-source tools like core-v-verif, verilator, and the GCC compiler to convert any C code into a binary memory initialization file. Additionally, the RISC-V toolchain provides synthesizable SystemVerilog RTL code, which loads a binary memory initialization file into the FPGA's internal BRAM. This binary memory initialization file contains the C code functionality, transformed into opcodes and commands for RISC-V.
\ifdefined\IEEE  \else    \vspace{-.250em}  \fi
    \item With the binary memory initialization file and the generated RTL code, the next step is synthesizing the design using the Xilinx Vivado tool. This process generates a full design bitstream, a partial design *.bit, and *.bin files. To obtain FAR addresses relevant to the logic elements in the design, we performed the bitstream generation twice—once with the per-frame CRC option (2a) disabled and once (2b) enabled\footnote{The per-frame CRC option explicitly writes the FAR address for each frame. This facilitates extracting FAR information for LUTs, which is not included in the *.ll file.}.
\ifdefined\IEEE  \else    \vspace{-.250em}  \fi
    \item We utilized the FAR addresses obtained in step 2b as input to the EPOCH's code, generating an executable *.elf file for the processing system.
\ifdefined\IEEE  \else    \vspace{-.250em}  \fi
    \item We programmed the SoC board with the design bitstream and EPOCH's executable *.elf file and monitored the application's execution on the device through the USB-to-serial interface. This interactive interface allows users to monitor, pause, preserve, and restore their designs on-demand.
\ifdefined\IEEE  \else    \vspace{-.250em}  \fi
\end{enumerate}

Table~\ref{tab:Complex_benchmar_resource} provides a comprehensive overview of resource utilization, latency, and state preservation/restoration times for diverse, complex benchmarks. Our experiments utilized a single partition, spanning two clock regions (X$0$Y$0$ and X$1$Y$0$) on the PL side, encompassing $10$K LUTs, $20$K FFs, $40$ BRAMs, and $60$ DSP slices, as depicted in Figure~\ref{fig:FPGA-Layout} (b).  We constrained the logic placement within this P-block to simplify the analysis. The P-block was configured with the \texttt{SNAPPING\_MODE}\footnote{\texttt{SNAPPING\_MODE} property when applied on P-blocks, adjusts the size and shape automatically.} property set to \texttt{ROUTING}. For $7$-Series designs, this is recommended when P-block(s) cover non-reconfigurable site types, such as IOB, configurations, or clocking columns~\cite{Vivado_Guide}. As described in the Vivado design reference guide, this setting impacts signal routing and resource utilization. We attribute the slight variations in LUTs and FFs utilization across all benchmarks in Table~\ref{tab:Complex_benchmar_resource} to this effect. The results demonstrate successful state save-and-restore operations across all benchmarks. 
\\\indent
Table~\ref{tab:Complex_benchmar_resource} also includes the duration required to pause and preserve the state of each benchmark, along with the time needed for state restoration, measured in milliseconds. Notably, the state preservation time directly correlates with the number of resources utilized by each benchmark. Given the utilization of a PCAP-readback mechanism, the FAR addresses as an `input' is necessary to read out and save the state of each logic element within each frame. Our experiments reveal that utilizing the PCAP interface at $50 $MHz, the preservation and restoration of a single frame took $62.2 $µs and $67.4$µs, respectively. The maximum frequency supported by the PCAP is $100$ Mhz. Through experimentation, we discovered that reading consecutive frames at $100$ MHz frequently caused the reconfiguration fabric to freeze. As a result, we conducted our investigations at $50$ MHz and incorporated additional delays during back-to-back frame readbacks. Theoretically, by eliminating these delays and operating the PCAP interface at $100$ MHz, the time required for state preservation and restoration could be reduced by at least half.
\\\indent
While some benchmarks in Table~\ref{tab:Complex_benchmar_resource} exhibit microsecond latencies, EPOCH currently operates with context-saving latencies in milliseconds. However, as the first out-of-the-box preemption solution, EPOCH aims to demonstrate its flexibility and broad applicability across various scenarios. It aims to be the first complete system enabling preemption in cloud-based FPGAs. The system's effectiveness depends on several factors: application demands, priority settings, and scheduling intervals. The lack of standardized datasets and benchmarks for preemption hinders the full quantification of these benefits. Establishing such datasets and benchmarks is critical yet often overlooked in FPGA scheduling research. Now that we have established the first \textit{practical} preemption framework with EPOCH, future efforts can concentrate on the issues that need in-depth study to establish a robust foundation for future investigations.
\section{Cautions and Usage} \label{sec: Use-cases}
\ifdefined\IEEE  \vspace{0em} \else \vspace{-.75em} \fi
In complex designs with multiple clock domains, interruptions can lead to problems due to interconnected clock domains running in a feedback \textit{loop}-like fashion. This \textit{loop} is disrupted when a design is paused and restarted at arbitrary clock ticks. To mitigate these risks, it is crucial to maintain the clock-tick relation between different clock domains. This can be achieved by inserting a handshake mechanism (transmit-receive acknowledgment) between the PS and the PL, driven by the slowest clock in the system. This handshake helps the PS identify optimal moments for halting the clock without compromising functionality but implicates design-time changes.
\\\indent Even if the restoration procedure is executed adequately, it is possible that the correct values of the FF and BRAM may not be accurately reflected in the running configuration. In some cases, this issue may lead to a design freeze.
Two approaches can be employed to address this problem. First, before restoring the design configuration, a blanking bitstream can be written that initializes the PL fabric resources to known static values. This helps in avoiding timing hazards during the pause-and-restore operations. However, this approach impacts performance by incurring a time overhead.

The second option is to enable the \texttt{RAR} option in the constraint file, which acts similarly to the bitstream blanking but incurs no time penalty. Activating RAR increases the partial-bitstream size due to the addition of the \texttt{Global-RESTORE} command and CLB-CFG type frames as described in Section~\ref{Architecture}. The \texttt{Global-RESTORE} initializes the FF's content safely and avoids hazards during PR operations. However, this must be triggered by asserting the \texttt{GSR} signal once during system initialization and after each state recovery operation to make it successful. Further details can be found in~\cite{Config_guide}.

\ifdefined\IEEE  \vspace{0em} \else \vspace{-1em} \fi
\section{Discussions}\label{sec: Discussions}
\ifdefined\IEEE  \vspace{0em} \else \vspace{-0.75em} \fi
This section describes the scope of the proposed work. In addition, some of the obstacles and practical constraints pertaining to this effort are examined.
\ifdefined\IEEE  \vspace{0em} \else \vspace{-1em} \fi
\subsection{Scalability}\label{Scalability}
\ifdefined\IEEE  \vspace{0em} \else \vspace{-.5em} \fi
In a cloud system with `N' applications/tenants, accommodating the additional `N+1' application poses scalability challenges. EPOCH effectively addresses such scalability needs through the four steps outlined in Figure~\ref{fig:EPOCH-Process-Flow}. Currently, EPOCH supports the \textit{design-time} generation of memory initialization files and bitstream generation (steps 1 and 2 in Figure~\ref{fig:EPOCH-Process-Flow}). In the event of a new application's arrival, the cloud vendor must reperform RISC-V compilation, which in turn updates the memory initialization files residing in FPGA BRAM. However, the Xilinx SoC allows flexible BRAM content updates from the PS side, which can help transform steps 1 and 2 from design-time to \textit{run-time}\footnote{Currently, EPOCH does not support run-time BRAM content update. However, implementing such support is straightforward and will be addressed as a future enhancement.}. Step 3 in Figure~\ref{fig:EPOCH-Process-Flow} is required only once to convey the logical mapping of the FPGA elements currently used to EPOCH. Since new tenants reuse existing FPGA slots, this step is unnecessary at runtime. Step 4 ensures the availability of the partial bitstream relevant to the application for the cloud vendor. While pre-generating these partial bitstreams for known applications is possible at design time, EPOCH also supports the addition of new partial bitstreams at \textit{runtime}.  

\ifdefined\IEEE  \vspace{0em} \else \vspace{-1em} \fi
\subsection{Device Migration}\label{Device-Migration}
\ifdefined\IEEE  \vspace{0em} \else \vspace{-0.5em} \fi
EPOCH, designed for $7$-Series FPGAs, recognizes the widespread use of UltraScale FPGAs in multi-tenant cloud systems. To integrate EPOCH with the UltraScale FPGA family, slight modifications will be required. Significantly, the number of \texttt{words} per frame in UltraScale FPGAs increases marginally from $101$ to $123$. Hence, the readback capture sequence and the suggested FPGA frame write sequence described in Table~\ref{tab:Frame_Read_Template} and Table~\ref{tab:Frame_Write_Template} will require minor modification. For readers looking for detailed and comprehensive information, we refer them to UltraScale configuration guide~\cite{tapp2015configuration}. Additionally, state capture and recovery time may increase with $60$ configurable logic blocks (CLBs) per frame, compared to $50$ CLBs per frame for $7$-Series FPGAs. This is contingent on the operating clock frequency of the specific UltraScale FPGA.

\ifdefined\IEEE  \vspace{-.75em} \else \vspace{-0.75em} \fi
\subsection{Comprehensive Benchmarking}\label{Comprehensive-Benchmarking}
\ifdefined\IEEE  \vspace{0em} \else \vspace{-0.5em} \fi
Existing benchmarks do not adequately address the critical requirement for preemption support due to their diverse objectives~\cite{Chstone, Rosetta, Rosetta2, MLSBench, jamieson2005verilog, hansen1999unveiling, das2006improved }. Due to the lack of preemption-supported benchmarks in existing literature and online resources, we were motivated to develop custom benchmarks that establish preemption in multi-cloud FPGAs. Acknowledging the limited availability of benchmarks that accommodate this crucial requirement, we argue that creating benchmarks that enable preemption is worthwhile. This unfilled requirement emphasizes the significance of conducting collaborative investigations and creating a complete set of benchmarks to evaluate and verify preemption capabilities in multi-tenant computing environments extensively.

\ifdefined\IEEE  \vspace{0em} \else \vspace{-0.75em} \fi
\subsection{Design Enhancement}\label{Design-Enhancement}
\ifdefined\IEEE  \vspace{0em} \else \vspace{-0.5em} \fi
EPOCH relies on the PCAP interface for its state capture and recovery procedures, a choice made to simplify design and reduce user intervention. While EPOCH currently supports PCAP, it can be extended to accommodate the ICAP interface. It's crucial to note that delivering partial bitstreams via ICAP involves different formatting than PCAP, specifically concerning endianness and bit-level swapping of the data within the bitstream~\cite{UG909}.

Prior works exploring overclocking the ICAP interface for accelerated partial bitstream delivery are available for readers interested in further optimizing EPOCH's performance~\cite{ICAP_Accelerate_1, ICAP_Accelerate_2, ICAP_Accelerate_3, ICAP_Accelerate_4, ICAP_Accelerate_5, ICAP_Accelerate_6, ICAP_Accelerate_7, VRZYCAP}. Xilinx recently introduced a fast partial reconfiguration mechanism that leverages DMA with ICAP, supporting $32$-bit wide bitstream data at $200$ MHz ($800$ MB/s) for monolithic devices~\cite{XAPP1338}. However, it's essential to be aware that opting for ICAP instead of PCAP may increase area requirements and design complexity, particularly for novice FPGA developers.


\ifdefined\IEEE  \vspace{0em} \else \vspace{-0.75em} \fi
\subsection{Multi-Slot Extension}\label{Multi-Slot-Design}
\ifdefined\IEEE  \vspace{0em} \else \vspace{-0.5em} \fi
EPOCH's prowess has been rigorously examined through meticulous testing, focusing on single and dual-slot designs.  In particular, the scalability of EPOCH's architecture guarantees that its implementation is not limited to a fixed quantity of slots; instead, it is designed to accommodate multiple slots. By strategically allocating distinct memory regions within the DRAM, EPOCH effectively captures the state information related to each FPGA slot. The allocation is carefully calculated to be proportionate to each slot's particular demands (underlying logic elements), thus enabling smooth and effective acquisition of their respective states. The flexibility exhibited during our testing highlights the adaptability and scalability of EPOCH in accommodating a wide range of FPGA configurations and multiple FPGA layout positions.
\ifdefined\IEEE  \vspace{0em} \else \vspace{-0.75em} \fi
\subsection{Logic and Memory Isolation}\label{Tenant-Isolation}
\ifdefined\IEEE  \vspace{0em} \else \vspace{-0.5em} \fi
EPOCH employs PR to partition the FPGA into distinct slots, each assigned to a unique tenant. This exclusive assignment ensures that the logic associated with each tenant is \textit{isolated} to its designated slot. By leveraging the PR capabilities of FPGAs in conjunction with the template provided in Table~\ref{tab:Frame_Write_Template}, EPOCH can dynamically preserve, preempt, and restore the context of particular regions without impacting the overall design. Memory isolation is achieved by assigning dedicated memory regions to each partition, preventing data interference. Furthermore, features like encryption and authentication can be integrated to safeguard the integrity and confidentiality of each tenant's design. However, memory content security and secure memory/data access require orthogonal efforts, which are currently out of the scope of this work~\cite{DRAM_Security_1, DRAM_Security_2, DRAM_Security_3, malik2020isolation,malik2024enabling,malik2025craft,malik2022Obfuscation}.
\ifdefined\IEEE \vspace{0em} \else \vspace{-1em} \fi
\section{Conclusions}\label{sec: conclusions}
\ifdefined\IEEE      \vspace{0em} \else     \vspace{-0em} \fi
Incorporating multi-tenancy in cloud FPGAs presents a compelling opportunity to enhance efficiency and economic viability. Current FPGA vendors do not offer dynamic preemption support for temporal/spatial tenants despite its acknowledged benefits. Moreover, conventional preemption mechanisms utilized in multi-CPU/GPU applications do not readily extend to cloud FPGAs due to the heterogeneous nature of FPGAs. 

This work presented EPOCH, a PCAP-based preemption framework that realizes cloud tenants' stateful preservation and recovery. EPOCH incurs zero-area overhead on the reconfiguration fabric, allowing for effortless integration with \textit{any} cloud application. Through comprehensive benchmarking using RISC-V softcore on a Xilinx SoC, we have demonstrated the generality and efficacy of our methodology. 

\ifdefined\IEEE       \vspace{0em} \else \vspace{1em} \fi


\bibliographystyle{IEEEtran}
\bibliography{References}

\end{document}